\documentclass[sigconf]{acmart}

	\usepackage{popets}
	\setcopyright{popets}
	\copyrightyear{YYYY}
	\acmYear{YYYY}
	\acmVolume{YYYY}
	\acmNumber{X}
	\acmDOI{XXXXXXX.XXXXXXX}
	\acmISBN{}
	\acmConference{Proceedings on Privacy Enhancing Technologies}
\newcommand{\eg}{e.g., }
\newcommand{\etc}{etc.}

\newcommand{\etal}{et al.\@\xspace}

\usepackage[utf8]{inputenc}

\usepackage{array}
\newcolumntype{L}[1]{>{\raggedright\let\newline\\\arraybackslash\hspace{0pt}}m{#1}}
\newcolumntype{C}[1]{>{\centering\let\newline\\\arraybackslash\hspace{0pt}}m{#1}}
\newcolumntype{R}[1]{>{\raggedleft\let\newline\\\arraybackslash\hspace{0pt}}m{#1}}

\usepackage{array}
\newcolumntype{P}[1]{>{\raggedleft\arraybackslash}p{#1}}

\newcommand{\postquote}[1]{{\it``#1''}}

\newcommand{\blockpostquote}[1]{{\begin{quote}{\it``#1''}\end{quote}}}

\newcommand{\subreddit}[1]{{\it #1}}

\usepackage{seqsplit}

\usepackage{hyperref}       %
\usepackage{url}
\usepackage{xspace}
\usepackage{graphicx}
\usepackage{enumitem}
\usepackage{booktabs}
\usepackage{multirow}
\usepackage{amsfonts,amsthm,amsmath}
\usepackage{subcaption}
\usepackage{latexsym}       %
\usepackage{wasysym}        %
\usepackage{tcolorbox}
\tcbuselibrary{breakable}
\usepackage{tikz}
\usepackage{xurl}
\usepackage{ellipsis}

\usepackage{xcolor}
\usepackage{framed}
\usepackage{bm}
\usepackage{tabularx}
\usepackage{pifont}

\newcommand{\sarah}[1]{}
\newcommand{\kurt}[1]{}
\newcommand{\sai}[1]{}
\newcommand{\tara}[1]{}
\newcommand{\nina}[1]{}
\newcommand{\sunny}[1]{}
\newcommand{\animesh}[1]{}
\newcommand{\todo}[1]{}
\newcommand{\pgk}[1]{}

\newcommand{\pss}{digital privacy, safety, or security\xspace}
\newcommand{\pssand}{digital privacy, safety, and security\xspace}

\hypersetup{
    colorlinks=true, %
    linkcolor=red, %
    citecolor=blue, %
    filecolor=magenta, %
    urlcolor=purple, %
    linkcolor=magenta %
}

\begin{document}

\title{Understanding Help Seeking for Digital Privacy, Safety, and Security on Reddit}

\author{Kurt Thomas}
\orcid{0000-0002-3762-5851}
\email{kurtthomas@google.com}
\affiliation{%
  \institution{Google}
  \city{} %
  \state{} %
  \country{} %
}

\author{Sai Teja Peddinti}
\orcid{0009-0007-3242-9353}
\email{psaiteja@google.com}
\affiliation{%
  \institution{Google}
  \city{} %
  \state{} %
  \country{} %
}

\author{Sarah Meiklejohn}
\orcid{0000-0001-5671-1395}
\email{s.meiklejohn@ucl.ac.uk}
\affiliation{%
  \institution{University College London}
  \city{} %
  \state{} %
  \country{} %
}

\author{Tara Matthews}
\orcid{0009-0008-4805-8339}
\email{taramatthews@google.com}
\affiliation{%
  \institution{Google}
  \city{} %
  \state{} %
  \country{} %
}

\author{Amelia Hassoun}
\orcid{0000-0002-6328-4763}
\email{miahassoun@google.com}
\affiliation{%
  \institution{Google}
  \city{} %
  \state{} %
  \country{} %
}

\author{Animesh Srivastava}
\orcid{0009-0005-5439-4150}
\email{sranimesh@google.com}
\affiliation{%
  \institution{Google}
  \city{} %
  \state{} %
  \country{} %
}

\author{Jessica McClearn}
\orcid{0000-0002-1609-5257}
\email{Jessica.Mcclearn.2021@live.rhul.ac.uk}
\affiliation{%
  \institution{Google}
  \city{} %
  \state{} %
  \country{} %
}
\additionalaffiliation{%
  \institution{Royal Holloway, University of London}
  \city{}
  \state{}
  \country{}
}

\author{Patrick Gage Kelley}
\orcid{0000-0003-4405-0010}
\email{patrickgage@google.com}
\affiliation{%
  \institution{Google}
  \city{} %
  \state{} %
  \country{} %
}

\author{Sunny Consolvo}
\orcid{0000-0001-5337-328X}
\email{sconsolvo@google.com}
\affiliation{%
  \institution{Google}
  \city{} %
  \state{} %
  \country{} %
}

\author{Nina Taft}
\orcid{0009-0008-8450-9627}
\email{ninataft@google.com}
\affiliation{%
  \institution{Google}
  \city{} %
  \state{} %
  \country{} %
}

\begin{abstract}
The complexity of navigating digital privacy, safety, and security threats often falls directly on users. This leads to users seeking help from family and peers, platforms and advice guides, dedicated communities, and even large language models (LLMs). As a precursor to improving resources across this ecosystem, our community needs to understand what help seeking looks like in the wild. To that end, we build a scaled analysis pipeline leveraging LLMs to sift through over two billion Reddit posts from the last five years to identify where and for what users seek digital privacy, safety, or security help. We isolate four million relevant posts with 93\% precision and recall and automatically annotate each with the topics discussed (\eg privacy configurations, data misuse, account compromise, and more). We use this dataset to understand the scope and scale of help seeking on Reddit, the subcommunities that provide help, and to qualitatively analyze the types of help sought. Our work informs the development of better resources for users (\eg user guides or LLM help-giving agents) while underscoring the inherent challenges of supporting users through complex combinations of threats, platforms, mitigations, context, and emotions.
\end{abstract}


\renewcommand{\shortauthors}{Thomas et al.}

\maketitle

\section{Introduction}

Internet users are confronted with a broad array of threats to their digital privacy, safety, and security. Attacks include unintended data collection by platforms~\cite{gross2005information,razaghpanah2018apps}, compromise and fraud conducted by cybercriminals~\cite{thomas2017data,anderson2013measuring}, and interpersonal abuse by close relations~\cite{thomas2021sok,freed2018stalker,chatterjee2018spyware}. The burden of navigating these threats often falls on users who may lack the requisite technical expertise, tools, or resources to adequately protect themselves. This ultimately leads users to seek help from family and peers~\cite{redmiles2016learned,rader2015identifying, rader2012stories,freed2025help}, platforms and advocacy groups~\cite{redmiles2016learned,redmiles2020comprehensive, redmiles2016think}, dedicated online communities~\cite{morris2010people,ellison2013calling}, and more recently, large language models (LLMs)~\cite{chen2023can}.

A precursor to
improving help-giving resources---no matter where users turn for support---is understanding \emph{what} help seeking for digital privacy, safety, or security looks like in the wild. Prior work has focused on qualitative investigations of help seeking for individual topics, such as harassment and sexual abuse~\cite{andalibi2016understanding, wei2024understanding}; ensuring a user's privacy when interacting with smart home devices~\cite{li2023s}; or emerging AI chatbots~\cite{ali2025understanding}. Additionally, a wealth of research has explored the quality and effectiveness of digital-safety advice available online~\cite{redmiles2016think, coopamootoo2023equal, hashmi2025understanding, nicholson2019if,munyendo2026wonder,redmiles2016learned,redmiles2020comprehensive, herbert2023world, fagan2016they,ion2015no, busse2019replication}. However, the scale of users seeking help, how quickly help-seeking needs shift over time, and what specific needs are currently unmet by existing resources remains unknown.

In this paper, we study help seeking for digital privacy, safety, and security at scale through the lens of Reddit. Our dataset includes over two billion Reddit posts authored by users over five years from January 2021--December 2025. We focus on Reddit as users publicly ask bespoke questions to the broader community~\cite{morris2010people, ellison2013calling}, often relying on the platform's anonymity when discussing potentially sensitive topics~\cite{pan2023we, andalibi2016understanding,peddinticosn2014}. Our investigation focuses on three research questions:

\begin{enumerate}[label={\bfseries RQ\arabic*:},leftmargin=1cm, topsep=0pt,itemsep=0ex,partopsep=1ex,parsep=1ex]

    \item What is the scope and scale of help seeking for digital privacy, safety, and security on Reddit? How have trends evolved over time?

    \item What  subreddits (\eg communities) do users turn to for help with digital privacy, safety, and security?
    
    \item What types of support do users request? How complex is it to meet users' bespoke needs?
        
\end{enumerate}

We answer these questions using a mixed methods approach that blends quantitative and qualitative research. We demonstrate that by fine-tuning a Gemini model that detects help seeking for digital privacy, safety, and security, we are able to surface four million help-seeking questions across more than 5,000 subreddits with 93\% precision and recall. Using a similar process, we automatically annotate each post with any of nine possible topics (\eg \emph{Account tools} for 2FA, password managers, and access issues; or \emph{Privacy tools} for Tor, VPNs, or ad blockers).

We find that help seeking for digital privacy, safety, and security has rapidly grown over the last two years, with over 60,000 questions posted every month at the end of 2025. Most users seek support for scams (29\%), accessing or protecting their accounts (21\%), and navigating privacy tools and settings (19\%), but we also observe a rise in users seeking help for harassment (11\%) and related tools for blocking or reporting abusers (6\%). Thousands of subreddits help users to navigate these threats, with some---like \subreddit{r/scams}---providing specialized support for specific threats, while others---like \subreddit{r/facebook} or \subreddit{r/coinbase}---help users navigate digital privacy, safety, or security concerns associated with specific platforms. Across topics and subreddits, users ask others to help them make sense of situations, to provide guidance for what to do next or technical troubleshooting, or to provide emotional reassurance to put them at ease.

Our results underscore the brittleness of one-size-fits-all approaches to digital privacy, safety, and security. Users experience complex combinations of threats, platforms, mitigations, context, and emotions that shape their support needs. These nuances going beyond what is feasible of advice guides that are distilled into best practices meant to be consumed by all general internet users or specific populations. We discuss the implications this has for subreddits that provide help, online resources and help centers, as well as help-giving agents built on LLMs. Furthermore, as researchers explore benchmarks to assess the \pss knowledge of models~\cite{prakash2024assessment, tihanyi2024cybermetric, liu2024cyberbench,jing2024secbench}, these benchmarks should reflect the rich, real-world needs identified in our dataset. In summary, this paper makes the following key contributions:

\begin{itemize}[leftmargin=*,itemsep=5pt,topsep=10pt]
    \item \textbf{Scalable measurement pipeline:} We design and validate an end-to-end automated pipeline that identifies help-seeking for digital privacy, safety, and security and categorizes the topics involved.
    
    \item \textbf{Longitudinal measurement of help seeking:} We present a five-year measurement of help seeking trends on Reddit, analyzing 4~million posts across 5,425 subreddits.

    \item \textbf{Qualitative taxonomy of user needs:} We characterize the unmet needs of users that leads them to seek help and share shortcomings of monolithic, one-size-fits-all advice.
    
    \item \textbf{Open-source golden dataset \& artifacts:} We publicly release our annotated golden datasets, topic codebooks, prompts, annotated dataset to enable future research on help seeking and improving support.\footnote{\url{https://github.com/google/helpseeking}}
\end{itemize}

\section{Related work}

\paragraph{Advice versus help seeking} A significant body of work has explored privacy, safety, and security advice that distills the complexity of digital safety into best practices such as ``enable two-factor authentication'' or ``keep software up-to-date''~\cite{ion2015no}. Prior studies have examined where different populations turn to for advice~\cite{redmiles2016think, coopamootoo2023equal, hashmi2025understanding, nicholson2019if,munyendo2026wonder,redmiles2016learned}, the quality of advice available from online guides~\cite{redmiles2020comprehensive}, why advice is challenging to follow or fraught with misconceptions~\cite{herbert2023world, fagan2016they}, and how advice differs between experts and non-experts~\cite{ion2015no, busse2019replication}. Because advice is designed to reach large audiences, the information provided is often restricted to a prioritized list meant for general internet users or specific populations, precluding bespoke advice tailored to a single user's exact needs.

Our study focuses on help seeking, which we show encapsulates a far more complex set of needs than advice alone---users seek to make sense of a privacy, safety, or security issue they have encountered; ask for step-by-step guidance on resolutions; and often seek emotional support to put them at ease. Where advice guides may recommend ``Use the Tor browser'', help-seeking questions in our dataset are much more nuanced, such as ``Does running your own Tor bridge mitigate timing attacks from government agencies, or does connecting to the same bridge every time make you less anonymous?'' These bespoke questions emerge due to the limited information present in advice guides or generally online. Communities such as Reddit offer a way for users to connect with (potential) experts in order to receive tailored information, making them a unique lens to study help seeking.

\paragraph{Scope and quality of online self-help resources} Users can currently turn to a diverse set of online resources for advice related to digital privacy, safety, and security. These include curated educational resources for youth~\cite{digital_citizenship, google_bia} and the general public~\cite{staysafeonline}; help guides for specific \pss practices~\cite{frontlinedefenders,effsecuritystarterpack,privacyguides,icac_ocepi}; and platform-specific help centers. Redmiles \etal assessed the utility of many of these resources and found that while the advice is accurate, online guides fail to prioritize which advice to follow, leading to an overwhelming volume of recommendations~\cite{redmiles2020comprehensive}. Even experts struggle to prioritize advice~\cite{securityadvice2017,wei2023there}. Users are thus left to  navigate a sea of recommendations without the necessary technical expertise to assess the tradeoffs of various protections~\cite{securityadvice2017}. Researchers recently explored whether LLMs can assist with this knowledge and discovery gap, but found that the most sophisticated models currently available fail to offer consistently accurate advice~\cite{stackoverflow_chatgpt,chen2023can}. Our study helps to inform the design of better resources by examining the threats, platforms, and contexts for which users currently seek help.

\paragraph{Help seeking via online communities}
Help seeking refers to the actions taken by individuals to obtain assistance to meet their needs, or seek personal or psychological support~\cite{barker2005young}. Multiple factors influence where people turn to for help for \pss, such as age, gender, and socio-economic status~\cite{redmiles2016learned, redmiles2020comprehensive, redmiles2016think}. Online communities---like Reddit---have emerged as an important outlet for help seeking, in part due to their low barrier of entry and reduced fear of stigmatization~\cite{freed2025help,pan2023we, andalibi2016understanding,peddinticosn2014}. Prior studies of help seeking via online communities have narrowly focused on individual topics, such as harassment and sexual abuse~\cite{andalibi2016understanding, wei2024understanding}, mental health questions~\cite{jamie-tiktok-questions}, or interpreting the privacy and safety attitudes of users based on conversations surrounding IoT devices~\cite{li2023s} or AI chatbots~\cite{ali2025understanding}. Methodologically, these prior works typically rely on manual or keyword-based `vertical' analyses of single topics or specific subreddits, or leverage LLMs strictly for text summarization~\cite{antisdel2025data}. Such approaches either fail to scale, are affected by typos/slang~\cite{protick2026llm}, or suffer from lower classification performance (e.g., $F_1 \approx 0.7$ in~\cite{guan2025classifying}). Our study departs from these limitations in both scale and methodology. We examine a much larger scope of \pss help seeking in terms of the topics studied, timeframe, and scale of data analyzed while achieving state-of-the-art accuracy.

\begin{figure*}[t]
    \centering
    \includegraphics[width=0.95\textwidth]{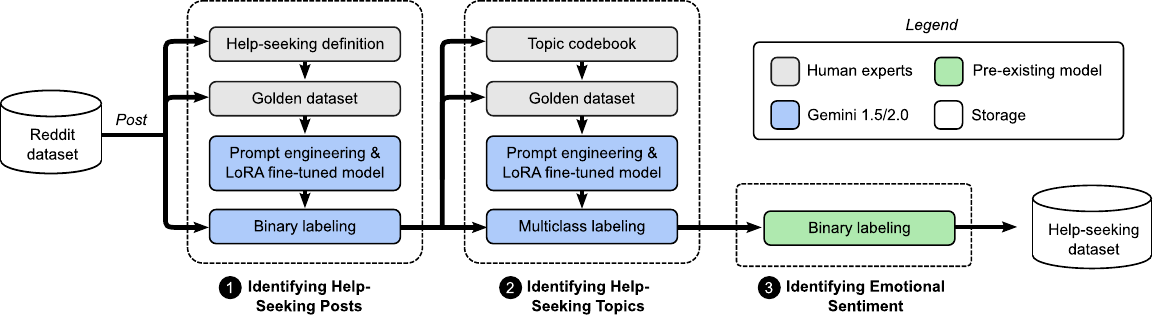}
    \caption{Automated pipeline for identifying posts seeking help for digital privacy, safety, or security. Human experts reviewed thousands of Reddit posts to define help seeking and to identify nine distinct help-seeking topics. These same experts annotated a golden dataset for fine-tuning which allowed an LLM to generalize and apply these definitions at-scale, in addition to isolating the emotion of the post.}
    \label{fig:helpseeking_llm_flow}
    \Description{A flowchart describing the analysis pipeline.}
\end{figure*}

\section{Methodology}\label{sec:methods}
We develop a pipeline that sifts through Reddit posts to identify help seeking for digital privacy, safety, or security. Our pipeline also automatically extracts the topics discussed (\eg privacy tools, scams) and the emotional sentiment of the post. We provide a high-level overview of our pipeline in Figure~\ref{fig:helpseeking_llm_flow}. We also discuss our qualitative analysis process which we apply to a sample of posts.

\subsection{Reddit dataset}
Our Reddit dataset is sourced from a pre-existing Reddit-compliant web crawl that respects its \texttt{robots.txt} and other rules for crawlers. The dataset consists of over 2 billion \emph{original posts}---the first post in a discussion thread. For simplicity, we refer to original posts hereafter as \emph{posts}. The periodic nature of crawling precludes collecting all \emph{comments} in the discussion thread or knowing when discussions conclude, so we omit all such comments from analysis. Our corpus spans five years, from January 1, 2021 to December 31, 2025. For each post, we have the text of the post, the username that authored the post, the timestamp when the post was created, the subreddit (\eg community) where the post appeared, and a unique post identifier. Our Reddit dataset contains posts from 8 million subreddits. For the purposes of our study, we restrict our analysis to (1) posts with at least 100 characters (77\% of posts meet this criteria) and (2) subreddits with at least 1,000 posts (93\% of posts meet this criteria). Here, our intent is to identify well structured help-seeking questions posed to more active subreddits, while ensuring low-density communities are not systematically lost. These two criteria reduce our search space to 1.5 billion posts appearing on 117,390 subreddits.

\subsection{Identifying help-seeking posts}
\label{sec:identifyinghelpseeking}
Given a post from our Reddit dataset, the first stage of our pipeline identifies whether the post involves seeking help for digital privacy, safety, or security (Figure~\ref{fig:helpseeking_llm_flow}, \ding{182}). 

\paragraph{Defining help seeking} To arrive at a definition of help seeking,
nine members of our research team with expertises spanning privacy, safety, and security reviewed 800 posts randomly sampled from 53 manually identified subreddits that focused on privacy (\eg \subreddit{r/privacy}, \subreddit{r/privacyguides}), safety (\eg \subreddit{r/scams}, \subreddit{r/stalking}), security (\eg \subreddit{r/infosec}, \subreddit{r/security}), and general technology support (\eg \subreddit{r/techsupport}, \subreddit{r/ask}). See Appendix~\ref{sec:golden_dataset_sampling} for a full list of subreddits. Researchers regularly reviewed posts, engaged in discussion on nuanced boundaries, and collectively refined a shared definition of help seeking.

For a post to be in scope for our study, the post had to (1) seek to prevent, or be experiencing, or know someone who is experiencing a digital privacy, safety, or security concern; and (2) ask for advice on how to respond to the concern, which might include general sense making (\eg ``Why is this happening?''), technical advice (\eg ``How do I block this person?''), therapeutic advice (\eg ``Is feeling this way normal?''), relationship advice (\eg ``Can I trust them?''), or advice about external support (\eg ``Can the police help?''). We also considered implicit requests for help as in scope (\eg ``I'm freaking out''). Per this definition, the following two posts would be in scope:
\blockpostquote{Is this a scam? I have a package from Amazon out today, but the e-mail and address seem a bit fishy.}
\blockpostquote{I'm freaked out because my name is searchable on Spokeo and Im afraid someone will dox me. What can I do?}

Our definition treats the following as out of scope: purely informational or news posts; incidental mentions of technology; exclusively seeking educational resources; commercial questions where the entity experiencing a concern is a business; or help seeking related to how to intentionally cause harm. Per these exclusions, the following two posts would be out of scope due to being purely informational and educational, respectively:
\blockpostquote{Was shopping on Facebook Marketplace. Beware of [seller] \dots They're a known scammer}
\blockpostquote{I want to grow my skills in cybersecurity \dots What do you recommend?}

{\noindent}For the remainder of the paper, we refer to posts that meet our definition as \emph{help-seeking posts}.

\paragraph{Curating a golden dataset} The same researchers who engaged in defining help seeking manually labeled a total of 2,000 posts to act as a golden dataset. Creating this dataset via random sampling across all of Reddit would be prohibitively expensive for producing a sufficient number of positive help-seeking examples---most of Reddit is not dedicated to digital privacy, safety, or security. Instead, we labeled the 800 posts that we previously discussed, as well as 1,200 new posts. We sampled this latter set by leveraging existing privacy, safety, and security taxonomies~\cite{scalingwall2015,securityadvice2017,beals2015framework,thomas2021sok,wei2024understanding,harkous2022hark,akgul2024decade,solove2005taxonomy,wangandkobsa2009} to enumerate 79 concepts (\eg end-to-end encrypted messaging, account takeover, permission settings, \etc) and conducted a stratified search across Reddit for these concepts by prompting Gemini 1.5 Flash (Appendix~\ref{sec:golden_dataset_sampling}).\footnote{Throughout this work, we use the Gemini model due to available compute capacity, but other similarly sized closed-source or open-weight models that support bulk querying and LoRA fine tuning would be viable alternatives.} Additionally, we used a preliminary version of our help-seeking prompt %
(Appendix~\ref{sec:helpseeking_prompt})
to identify posts that \textit{might} meet our definition of help seeking. These searches increased the subreddits covered by our golden dataset to a total of 749 (from 53).

To label every post in our golden dataset, two or more researchers independently coded each post following a consensus coding approach~\cite{cascio2019consensus}. In the event of disagreement, the broader research team engaged in resolution sessions to arrive at a consensus on the most appropriate label.\footnote{Per our approach to resolving disagreement, we do not report inter-rater reliability.} Of the 2,000 posts analyzed, we labeled 747 as help seeking and 1,253 as not help seeking. We caution this golden dataset is not truly random, and may contain some biases. Our goal was to rely on multiple sampling techniques to minimize bias, while expanding our search to all of Reddit with the assistance of LLMs. We further discuss limitations of our approach in Section~\ref{subsec:limitations}. For subsequent evaluations, we split our golden dataset, using 75\% (1,500 posts) for training, 10\% for validation (200), and 15\% for testing (300).

\paragraph{Prompt engineering \& fine-tuning} We considered two strategies for instructing an LLM on how to apply our help-seeking definition at scale: in-context learning and fine-tuning using Low-Rank Adaptation (LoRA)~\cite{hu2021loralowrankadaptationlarge}, which has demonstrated on-par or better performance compared to full fine-tuning for training datasets of our scale. We iteratively experimented with multiple prompt versions that differed in the task type (prediction-only vs. chain-of-thought generation of a justification prior to the prediction, which requires more tokens and incurs a higher cost) and whether to include examples to reduce task ambiguity (zero-shot vs.\ few-shot). Based on performance, we settled on a prediction-only prompt that consisted of three parts: (1) a definition of help seeking as detailed earlier; (2) a set of exclusion criteria as detailed earlier; and (3) a static, few-shot list of examples containing both positive and negative examples of help seeking. See Appendix~\ref{sec:helpseeking_prompt} for our finalized prompt.

We evaluated the in-context learning capabilities of Gemini 1.5 Flash\footnote{We opted for Gemini Flash over Gemini Pro as the former required less compute resources, which was critical to scaling to all of Reddit. Gemini 1.5 was the most sophisticated model available for our use case at the time we began the first stage of our pipeline development.} with the aforementioned prompt on our test dataset of 300 posts, achieving an F1 score of 78.7\% as shown in Table~\ref{table:candidate_models}. We deemed the model's precision of 68.1\% to be too noisy for the purpose of downstream accurate measurements.

As an alternative, we LoRA fine-tuned a Gemini 1.5 Flash model using our training and validation datasets and the same in-context prompt.  We configured training to use a rank of 4; 4,000 training steps; a batch size of 2; and a learning rate of 0.0001. We report the results of this LoRA model on our test dataset of 300 posts in Table~\ref{table:candidate_models}. Overall, the fine-tuned model achieved an F1 score of 90.7\%, outperforming in-context learning by 12\% when evaluated on the same dataset.

\begin{table}[t]
    \caption{Performance of our LoRA fine-tuned Gemini 1.5 Flash model for identifying help-seeking posts, as well as a few-shot prompt using Gemini 1.5 Flash, based on holdout test sample of N=300.}
    \footnotesize
\centering
\begin{tabularx}{0.8\columnwidth}{X|r|r}
\toprule
   \bf Metric & \bf Fine-tuned model & \bf Few-shot prompt\\
   \midrule
   Recall & \bf 93.3\% & 93.3\% \\
   Precision & \bf 88.3\% & 68.1\%\\
   F1 & \bf 90.7\% & 78.7\%\\
   Accuracy & \bf  93.3\% & 84.4\%\\
   \bottomrule
\end{tabularx}

    \label{table:candidate_models}
\end{table}

\paragraph{Identifying help-seeking posts at-scale} Applying our fine-tuned model to all posts in our Reddit dataset would be prohibitively expensive. We instead relied on stratified sampling to first isolate relevant subreddits where people engaged in help seeking for digital privacy, safety, and security. We took a stratified sample of 100 posts from each of the 117,390 subreddits in our dataset totaling roughly 11.7 million posts, and labeled each post with our fine-tuned model. In total, 6,774 subreddits contained at least one post identified by our model as help seeking. We checked whether any of these subreddits are now private, quarantined, or banned\footnote{For example, \textit{r/tiktokhelp} was a popular help seeking subreddit, but was banned due to Reddit's rules requiring moderators.} compared to when they were crawled. As of January 7, 2026, 1,349 of the 6,774 relevant subreddits were inaccessible. We omitted these subreddits from subsequent analysis. 

Across the remaining 5,425 public, active, relevant subreddits, there were a total of 118 million posts. We classified all of these with our fine-tuned model, identifying 4 million help-seeking posts (3.4\%). We summarize our dataset in Table~\ref{table:final_dataset}. This dataset forms the basis of our subsequent topic labeling and analysis. 

\begin{table}[t]
    \caption{Summary of Reddit posts identified by our fine-tuned model as help seeking for digital privacy, safety, or security.}
    \footnotesize
\centering
\begin{tabularx}{0.9\columnwidth}{X|r}
\toprule
   \bf Dataset  & \bf Value\\
   \midrule
   Help-seeking posts & 4,078,631\\ 
   Unique subreddits & 5,425 \\
   Unique users & 2,714,850 \\
   \midrule
   Time period & January 1, 2021 -- December 31, 2025\\
   \bottomrule
\end{tabularx}

    \label{table:final_dataset}
\end{table}

\begin{table*}[t]
    \centering
    \caption{Digital privacy, safety, and security topics that emerged from our thematic analysis. The frequency values indicate their representation among the 750 hand-validated help-seeking posts. }
    \footnotesize
\begin{tabularx}{\textwidth}{r|X|r}
\toprule
\bf Topic & \bf Description & \bf Frequency\\
\midrule
\it Account tools &  Topics include account recovery (including cryptocurrency wallets), account security configurations (\eg 2FA, SMS authentication), forgotten passwords, password managers, or inaccessible accounts or devices. & 128 \\
\midrule
\it Safety tools & Topics include platform-provided safety tools such as blocking, reporting, takedown, and other moderation tools. & 63\\ 
\midrule
\it Security tools & Topics include popular security tools (\eg anti-virus, hardware cryptocurrency wallets), security practices (\eg encryption, access control, logging, firewalls), or general security topics (\eg jailbreaking, network security). & 125 \\
\midrule
\it Privacy tools & Topics include popular privacy tools (\eg Tor, VPNs, ad blockers), configuring privacy settings (\eg cookies, E2EE), data controls (\eg permissions, settings), or general privacy topics (\eg preventing tracking). & 152\\
\midrule
\it Compromise & Users who suspect or describe behavior that suggest their account or device may have been compromised. Topics include account hacking, suspicious sign-ins or OTPs, malware, or suspicious files. & 123\\
\midrule
\it Platform actions & Users experiencing friction with platform policies or safety barriers. Topics include being banned, suspended, or falsely reported; access being blocked to websites or services; or transactions being incorrectly flagged as fraud. & 99\\
\midrule
\it  Scams & Users mentioning fraud or scams; also transactions or missing funds or goods. Topics include fake job postings, sextortion, spam, unauthorized transactions, stolen funds (including cryptocurrencies), or suspicious webpages or links. & 173 \\
\midrule
\it Harassment & Users mentioning harassment or sexual abuse. Topics include toxic comments, stalking, doxxing, grooming, interpersonal abuse, or image-based sexual abuse (\eg cyberflashing, non-consensual explicit imagery). & 115\\
\midrule
\it Data concerns & Users concerned about their data being exposed, misused, or inappropriately collected. Topics include oversharing, breaches, retention, collection, and consent. & 101 \\
\bottomrule
\end{tabularx}

    \label{tab:taxonomy}
\end{table*}

\paragraph{Final validation} As a final validation step, we estimate the precision of all posts automatically labeled as help seeking in our dataset. Using a random sample of N=322 posts, two coders independently reviewed and manually labeled each post as help seeking or not before resolving any disagreement as before. This sample consisted of 300 help-seeking posts, and 22 false positives, which equates to a precision of 93.2\% ($\pm 2.76\%$ at 95\% CI).\footnote{This differs from our estimate in Table~\ref{table:candidate_models} in part due to our final dataset reflecting posts from 5,425 subreddits, compared to our golden test dataset sampled from 749 subreddits.} For estimating recall, it is infeasible to manually review a statistically significant volume of negative model outputs to identify false negatives. Instead, we estimate the recall of our model based on our golden dataset to be 93.3\% ($\pm$ 2.82\% at 95\% CI).

\subsection{Identifying help-seeking topics}
\label{sec:identifyingtopics}
Given a help-seeking post, the second stage of our analysis pipeline identifies the digital privacy, safety, or security topics discussed within the post (Figure~\ref{fig:helpseeking_llm_flow}, \ding{183}), with the potential of multiple labels applying to a single post. 

\paragraph{Enumerating help-seeking topics} 
We began with deductive thematic analysis of prior taxonomies~\cite{scalingwall2015,securityadvice2017,beals2015framework,thomas2021sok,wei2024understanding,harkous2022hark,akgul2024decade,solove2005taxonomy,wangandkobsa2009} to identify common \pss topics. We supplemented this with inductive, iterative coding of 300 posts (from the final validation dataset used to estimate precision in the wild) by nine researchers to refine the topics based on the data. Our hybrid approach~\cite{fereday2006inductive} yielded nine top-level themes---each containing multiple sub-themes. We refer to these hereafter simply as \emph{topics}. We list the nine topics and provide a short description of each in Table~\ref{tab:taxonomy}. For example, this post from \subreddit{r/fidelityinvestments} relates to the \textit{Scam} topic:

\blockpostquote{New Scam? Got a weird email claiming to be Fidelity. On a home I bought 2 years ago. They even got the address spelled incorrectly\dots}

{\noindent}Whereas the following post from \subreddit{r/1password} relates to both the \textit{Account Tools} and \textit{Compromise} topics:

\blockpostquote{Help! Master password may have been changed by a hacker. I am in a bit of a panic as my master password to open 1password is not working\dots}

{\noindent}See Appendix~\ref{sec:topic_codebook} for our full codebook.

\paragraph{Curating a golden dataset} The same researchers involved in thematic analysis also manually labeled a golden dataset of 750 posts across all nine topics, where multiple topics might apply. Here, we re-used the 300 posts that we previously discussed. We also used a preliminary version of our topic classification prompt (Appendix~\ref{sec:helpseeking_topic_prompt}) to conduct a stratified search across all 9 topics, yielding another 450 posts.
These additional posts were sampled to improve representation of harder-to-find topics and also to correct for common errors from the prompt. Two or more researchers independently coded each post for all nine topics, with any disagreement resolved by the broader research team. Our final topic golden dataset contains 6,750 labels---9 binary topic labels for 750 posts. Table~\ref{tab:taxonomy} provides a breakdown of the prevalence of each topic in our golden dataset. For subsequent experiments, we stratified and split the 6,750 labels, using 50\% for training, 15\% for validation, and 35\% for testing.

\paragraph{Prompt engineering \& fine-tuning} We again considered both in-context learning and LoRA fine-tuning to apply our multi-label topic codebook at-scale. For the overall prompt structure, we experimented with chain-of-thought, few-shot examples, exposing only a single topic definition, or all topic definitions at once. Ultimately, for in-context learning we settled on a ten-shot prompt with a single topic targeted per prompt. To select the few-shot examples, we created Gemini text embeddings of all training samples, selecting the 10 semantically closest neighbors for each test sample under evaluation. However, F1 performance of this approach ranged from 73\%--83\%, which we considered unacceptable.

As an alternative, we LoRA fine-tuned a Gemini 2.0 Flash model\footnote{At this time of our experimentation, Gemini 2.0 was now available. As before, we opted for Flash over Pro due to available compute.} using a prompt covering every topic, but with a switch statement to produce a single label per prompt call (Appendix~\ref{sec:helpseeking_topic_prompt}). We configured training to use rank=4; training steps=3000; batch size=16; and learning rate=0.0001. We report the F1 score per topic in Table~\ref{table:taxonomy_models} with an overall F1 score of 92.0\%, outperforming our in-context approach.

\paragraph{Identifying help-seeking topics at-scale} We applied our topic model to all 4 million help-seeking posts, resulting in 36 million inferred topic labels. 

\begin{table}[t]
    \caption{Performance of our LoRA fine-tuned Gemini 2.0 Flash model for identifying help-seeking topics, based on a holdout test sample of N=2,362 individual topic labels.}
    \footnotesize

\centering
\begin{tabularx}{0.95\columnwidth}{X|rrrr}
\toprule
\bf Topic & \bf Recall & \bf Precision & \bf F1 & \bf Accuracy \\
\midrule
Account tools & 95.1\% & 86.7\% & 90.7\% & 97.0\% \\
Safety tools & 95.7\% & 91.7\% & 93.6\% & 98.9\% \\
Security tools & 85.0\% & 91.9\% & 88.3\% & 96.6\% \\
Privacy tools & 96.2\% & 87.9\% & 91.9\% & 96.6\% \\
Compromise & 97.9\% & 95.8\% & 96.8\% & 98.9\% \\
Platform actions & 91.7\% & 97.1\% & 94.3\% & 98.5\% \\
Scams & 96.8\% & 90.9\% & 93.7\% & 97.0\% \\
Harassment & 95.2\% & 95.2\% & 95.2\% & 98.5\% \\
Data concerns & 95.0\% & 90.5\% & 92.7\% & 97.7\% \\
\midrule
Overall & 94.5\% & 89.6\% & 92.0\% & 97.3\% \\
\bottomrule
\end{tabularx}

    \label{table:taxonomy_models}
\end{table}

\subsection{Identifying emotional sentiment}

Prior work has demonstrated that it is feasible to accurately extract users' emotions from text~\cite{harkous2022hark,akgul2024decade}. To evaluate if certain help-seeking topics are associated with strong (positive or negative) emotions, we obtained the emotions classifier from Harkous \etal~\cite{harkous2022hark} and used it to identify the primary emotion across all of the 4 million posts in our help-seeking dataset. The classifier yields one of 28 emotions, with only a single label produced per post.

\subsection{Qualitative analysis of help seeking}
\label{sec:method-needs}
In addition to the qualitative methods we used to iteratively refine our taxonomy and label definitions, we qualitatively explored a random sample of help-seeking posts and topics to holistically understand the nature of support that users sought. We again relied on a hybrid deductive and inductive coding approach~\cite{fereday2006inductive}. We developed a codebook by starting with initial help-seeking codes from Wei et al.~\cite{wei2024understanding} (who studied help seeking for image-based sexual abuse), examining help-seeking posts, and inductively refining the codes to ensure they were grounded in the data. Two researchers then independently coded 250 randomly sampled help-seeking posts (which had previously been coded for \pss topics). They regularly met to resolve disagreements (following a consensus coding approach~\cite{cascio2019consensus}), interpret the coded data, and develop themes. By 250 posts, our codes and themes were stable and we ceased further analysis. 

Our final dataset of 250 posts includes a set of \textit{need type} codes (i.e., different kinds of help people indicated they need in their post)---such as a need to better understand their situation (\textit{sensemaking}); a need for advice on how to act given their situation (\textit{guidance}); or a need for reassurance, validation, or self expression in response to emotions (\textit{therapeutic}). It also includes a set of \textit{domain} codes (i.e., an area of knowledge or information) that were applied when knowledge of the domain would likely be necessary to interpret the request---such as \textit{technical} (relating to technology), \textit{platform} (a subset of \textit{technical} related to the use or behavior of specific platform UI flows, functions, procedures, or policies), or \textit{risk/harm} (relating to digital-safety risks, the potential for tech-facilitated harm by an attacker, or attackers’ behavior).
We also annotated the structure of each post,
by coding when \pss topics in a post represented: the \textit{cause} that initiated help seeking; 
\textit{practices} a user reported; and ultimately the \textit{question} posed by the user.

\subsection{Limitations}\label{subsec:limitations}
As with any quantitative and qualitative study, our methods have several limitations. Our crawl of 2 billion Reddit posts may be incomplete, resulting in some help-seeking posts being omitted from our study. While Reddit has over a billion monthly active users, its user base is skewed towards the United States and to younger people, which may bias the types of help seeking questions asked~\cite{reddit-mau}. Our reliance on Gemini 1.5 and 2.0 to identify help-seeking posts and topics incurs both false positives and negatives. Such errors may be non-uniform throughout time or may disproportionately apply to some help-seeking (sub-)topics more than others. While we provide estimates for the precision and recall of all our classifiers, our golden datasets may be biased, leading to analysis errors. Due to the potentially compounding factors of multiple classifier errors, we omit any discussion of p-values from our study. Finally, our study is limited to what users posted to Reddit at the start of a thread (\eg original post). Self-censorship, community rules, and platform reach may influence which users and what topics appear in our dataset. Likewise, any follow-up details posted in nested threads are not part of our analysis.

\section{Help-seeking trends}\label{sec:topic_trends}
We begin our results by examining help seeking for digital privacy, safety, and security at-scale. This includes measuring how the volume of help seeking has changed over time, identifying the topics for which users sought help, and exploring the relationship between help-seeking topics and  emotions.

\begin{figure}[t]
    \centering
    \includegraphics[width=0.95\columnwidth]{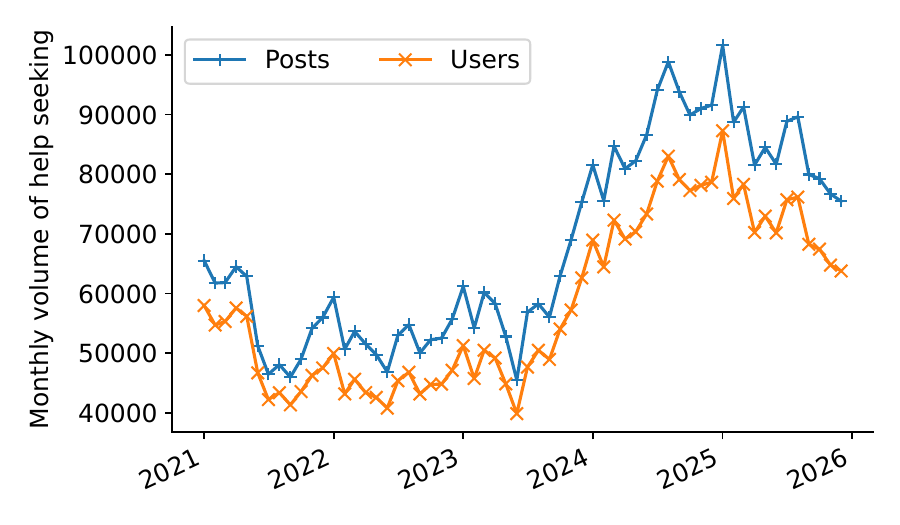}
    \caption{Volume of monthly help-seeking activity on Reddit for digital privacy, safety, and security concerns. (Note truncated y-axis.)}
    \label{fig:monthly_help_seeking}
    \Description{A temporal plot showing growth of help-seeking activity.}
\end{figure}

\paragraph{Growth of help-seeking posts over time} 
Overall, our pipeline identified 4 million help-seeking posts on Reddit from January 2021--December 2025. Figure~\ref{fig:monthly_help_seeking} shows how help seeking grew as a general practice during this time. While relatively stable from 2021--2023, the volume of posts and unique monthly users (\eg the authors of posts) accelerated in 2024 before declining in late 2025. The peak in January 2025 reflects over 102,000 posts in a single month from 87,000 unique users. 
While there is no clear explanation for the decline in 2025, we speculate the rise in popularity of LLMs may have reduced the volume of help seeking questions posed to Reddit.

\begin{table}[t]
    \centering
    \caption{Breakdown of help-seeking topics by volume and as a percentage of all help-seeking posts. Totals do not add to 100\% as multiple topics could apply to a single post.}
    \footnotesize

\centering
\begin{tabularx}{0.45\textwidth}{X|c|c}
\toprule
\bf Topic & \bf Volume & \bf Percentage \\
\midrule
Scams & 1,188K & 29.1\% \\
Account tools & 839K & 20.6\% \\
Privacy Tools & 779K & 19.1\% \\
Data concerns & 649K & 15.9\% \\
Compromise & 642K & 15.7\% \\
Security Tools & 533K & 13.1\% \\
Platform actions & 515K & 12.6\% \\
Harassment & 442K & 10.8\% \\
Safety tools & 243K & 6.0\% \\
\bottomrule
\end{tabularx}

    \label{table:topic_distribution}
\end{table}

\paragraph{Breakdown of help-seeking topics}
Table~\ref{table:topic_distribution} shows the breakdown of help-seeking topics in our dataset. Scams represented one of the largest sources of help-seeking posts (29\%). Relevant posts touched on having fallen victim to scams (\postquote{I gave personal details to a scammer\dots what else can I do?}), identifying suspicious messages and websites (\postquote{Is this a legit email? I've gotten about 6 of these recently.}), unexplained charges (\postquote{I am wondering if anyone else has seen a similar charge?}), missing funds (\postquote{Accidentally sent my life savings to a BTC address}), and more.

Account tools (21\%) was also a popular focus for help seeking. Relevant posts touched on login issues (\postquote{Why can't I login?}), uncertainty with security settings (\postquote{I have set up a 2FA \dots however [hacker is] still able to log in.}), and how to use various account security mechanisms like security questions, password managers, passkeys, and more. Privacy tools (19\%) was another popular topic. Here, users asked about popular privacy tools like Tor, VPNs, and ad blockers (\postquote{Does AdGuard VPN break Google functions?}), general privacy practices (\postquote{How do I get internet on my laptop away from home privately?}), and configuring privacy settings (\postquote{Is there a one click way to add the current website to the `Delete cookie and site data on restart' }).

Harassment (11\%) and Safety tools (6\%) were less common compared to other help-seeking topics. Users discussed various forms of interpersonal abuse from ex-partners (\postquote{I think my ex-partner is controlling what I see on social media}), family members (\postquote{my mother was checking through my laptop}), and strangers encountered online (\postquote{Had some kind of raider join me and my friend's server}). Safety tools adjacent to these topics included discussions of blocking people (\postquote{Why can a blocked caller still text me?}), reporting accounts, communities, or content (\postquote{Do reports actually do anything?}), or having accounts or content taken down (\postquote{Can you help me getting his account down?}).

We caution interpreting these metrics as a prioritized ranking due to how model errors in topic detection may influence our estimates (discussed previously in Table~\ref{table:taxonomy_models}). Furthermore, prevalence is just one metric: severity and actionability also are critical dimensions to consider when planning interventions~\cite{wei2023there, scheuerman2021framework}.

\begin{figure}[t]
    \centering
    \includegraphics[width=\columnwidth]{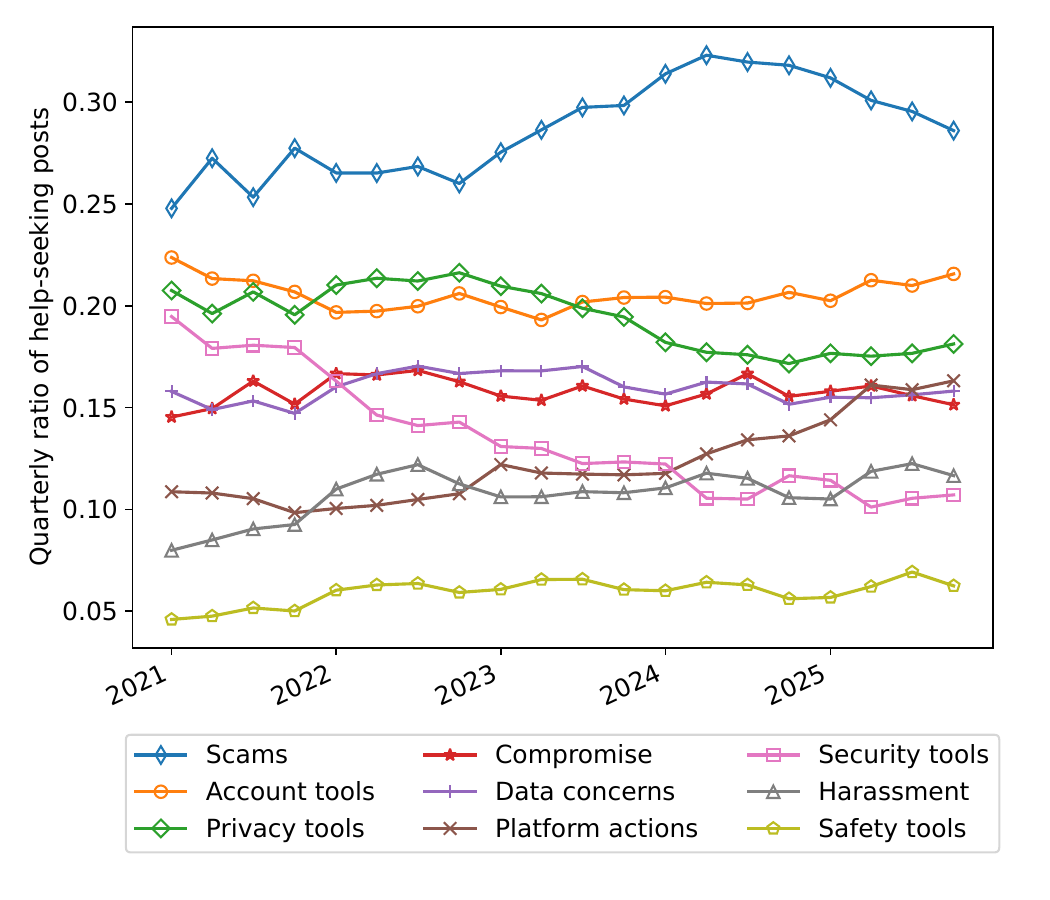} \vspace{-20pt}
    \caption{Temporal trends in help-seeking topics. Each line represents the ratio of posts that contained a topic versus all help-seeking posts, per quarter.}
    \label{fig:monthly_taxonomy_ratio}
    \Description{A temporal plot showing the trends in different help-seeking topics.}
\end{figure}

\paragraph{Evolution of help-seeking topics over time} Figure~\ref{fig:monthly_taxonomy_ratio} shows the relative growth and decline of help-seeking topics across posts, controlling for any quarterly changes in the  number of help-seeking posts overall. The ratio of posts related to Scams steadily increased over time from 24\% of posts at the start of 2021 to 32\% of posts in 2024, before declining to 29\%. By contrast, Compromise---which covers topics related to account hijacking, suspicious sign-ins, infected devices, and more---has remained relatively stable at 15--17\% of posts. Security tools---such as anti-virus, firewalls, encryption, hardware wallets, and more---has steadily declined from a peak of 19\% of posts in early 2021 to 10\% of posts at the end of 2025.\footnote{While the proportion of posts related to Security tools has decreased, we do not observe a drop in the absolute volume of posts related to this topic---other topics are merely growing faster.} Smaller shifts are also apparent, such as friction with Platform actions---covering account suspension or having transactions blocked---modestly increasing from a low of 10\% posts to now 16\% of posts in 2025. Monitoring these dynamics (along with the overall volume of help seeking) can help researchers and practitioners to identify emerging issues as well as longstanding unmet user needs. For example, our current data makes clear that Scams represent a sustained share of all \pssand help seeking on Reddit; users appear to be struggling to assess the authenticity of content they encounter online.

\begin{figure}[t]
    \centering
    \includegraphics[width=0.8\columnwidth]{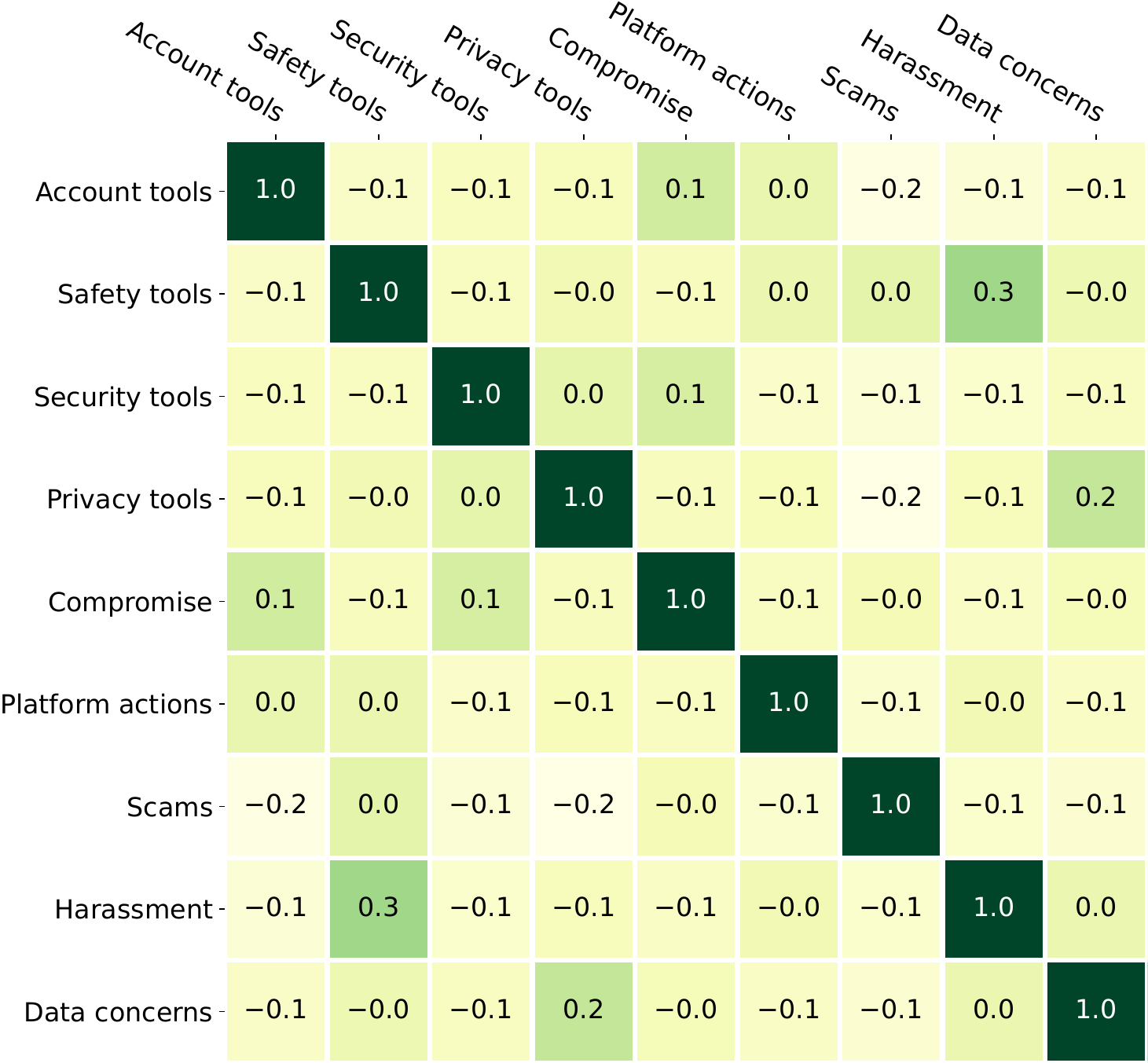}
    \caption{Heatmap showing Pearson correlations between each pair of help-seeking topics.}
    \label{fig:topic_correlation}
    \Description{A heatmap showing Pearson correlations between help-seeking topics.}
\end{figure}

\paragraph{Correlation between help-seeking topics} We calculate the Pearson correlation between help-seeking topics---where a single post can touch on multiple topics---to identify co-occurring topics. We find that help-seeking topics are largely independent, with only some weak correlations. For example, posts that touch on Compromise as a topic may also discuss Account tools (0.1) and Security tools (0.1) as prior practices or potential remedies. Similarly, some posts related to Harassment discuss Safety tools (0.3); and posts discussing Data concerns touch on Privacy tools (0.2). 
See Figure~\ref{fig:topic_correlation} for a full breakdown. The general lack of overlap between help-seeking topics within a single post points to the possibility of tailoring advice and tooling per topic, rather than designing a monolithic resource where users need to navigate every possible topic simultaneously.

\begin{figure}[t]
    \centering
    \includegraphics[width=\columnwidth]{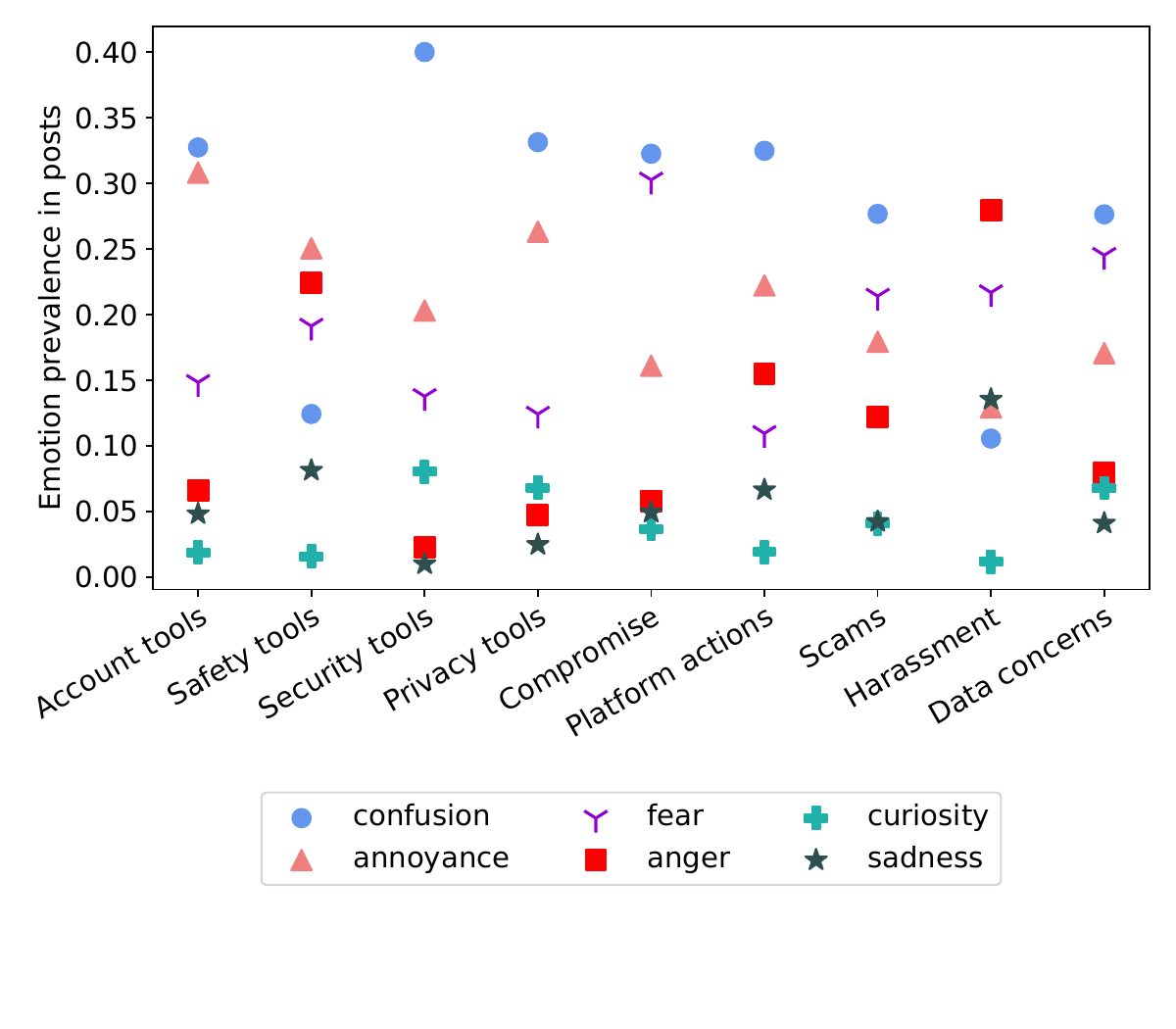}
    \vspace{-40pt}
    \caption{Top primary emotions identified in posts for each of the nine help-seeking topics.}
    \label{fig:topic_emotions}
    \Description{A scatter plot showing the distribution of top emotions across help-seeking topics.}
\end{figure}

\paragraph{Breakdown of primary emotions by help-seeking topic} Of 28 possible emotions, we found that just 6 primary emotions accounted for 88\% of posts: confusion (32\%), annoyance (22\%), fear (16\%), anger (9\%), curiosity (5\%), and sadness (4\%). A detailed breakdown of emotions by help-seeking topic is shown in Figure~\ref{fig:topic_emotions}. For an example of confusion, one user asked on \subreddit{r/paypal} about the risks of sharing their phone number with the platform:
\blockpostquote{I know this might sound silly, but if I use my phone number in the creation of an account, can they charge or do anything with my phone?}
 
Emotions offer an alternate perspective (in addition to prevalence and actionability discussed earlier) to consider when planning interventions. Prior frameworks also advocate that interventions should take into consideration the emotional state of a user~\cite{matthews2025tochi,scheuerman2021framework}.

\section{Help-seeking subreddits}
\label{sec:subreddits}
We continue our results by examining \textit{where} users seek help on Reddit. Here, we measure the most popular help-seeking subreddits and explore how they all orient around three community structures: general support; platform-specific support; and specialized support for harmful experiences, tools, and best practices.

\paragraph{Popular help-seeking subreddits} Table~\ref{tab:top-10-volume} shows a breakdown of the top 10 subreddits ranked by volume of help-seeking posts. We find \subreddit{r/scams} was the most popular subreddit, contributing 3.7\% of all help-seeking posts. That subreddit positions itself as a place ``for people wanting to educate themselves, find support, and discover ways to help a friend or loved one who may be a victim of a scam.'' According to Reddit's official statistics~\cite{reddit-top-communities}, this community had 1.3 million members as of January 2026 and was in the top 1\% of communities. Along with \subreddit{r/privacy}---which had 1.6M members---these subreddits provided what we call \emph{specialized support} for help-seeking questions. Among the top 10 subreddits, these two had the highest \emph{density} ($>$57\%) of help-seeking posts---calculated as the fraction of all posts on a subreddit that asked for help specifically about digital privacy, safety, or security.

Five subreddits in the top 10 provided what we call \emph{general support} (\subreddit{r/techsupport}, \subreddit{r/legaladvice}, \subreddit{r/advice}, \subreddit{r/help}, and \subreddit{r/nostupid-questions}). For example, \subreddit{r/techsupport} advertises itself as a community for anyone ``stumped on a tech problem.'' Illustrative questions from our dataset touched on \postquote{troubles with proxy settings}, \postquote{receiving [unwanted] random links}, worry after \postquote{browsing with my firewall turned off}, or why a file was \postquote{considered malicious in Virus Total}.
Despite densities of only 2--26\%, these subreddits likely served as easily-discoverable starting places even if more specialized subreddits existed. 

Three subreddits in the top 10 focused on what we call \emph{platform support} (\subreddit{r/facebook}, \subreddit{r/instagram}, \subreddit{r/coinbase}). One user on \subreddit{r/facebook} posted 
\postquote{someone hacked my account and I can't reclaim it}. Another user on \subreddit{r/coinbase} had their account \postquote{restricted from selling, withdrawing, and sending crypto} and was looking for solutions. These users may have been seeking platform-specific remedies or information, a phenomenon we explore in Section~\ref{sec:qual_analysis}. The relatively high densities for these subreddits (37-50\%) suggest that communities can operate as unofficial, quasi-support channels for privacy, safety, and security concerns for platforms.

Beyond the top 10 subreddits---which contributed 17\% of help-seeking posts---the cumulative volume of help-seeking posts followed a log-scale distribution. The top 100 subreddits contributed 46\% of posts, the top 1,000 contributed 88\% of posts, before a long tail of 5,425 total subreddits. This long tail illustrates that users seek help for \pssand across thousands of communities, with no single subreddit being dominant.

\begin{table}[t]
\centering
\caption{Top 10 subreddits ranked by the \textit{number of help-seeking posts}. We report the overall volume of help-seeking posts per subreddit, the overall percentage of help-seeking posts the subreddit contributes, and the fraction of all posts within a subreddit classified as help seeking for \pss which we refer to as \emph{density}.}
\footnotesize
\begin{tabularx}{0.8\columnwidth}{Xrrr}
\toprule
\bf Subreddit & \bf{Volume} & \bf{Percentage} &\bf{Density} \\
\midrule
r/scams & 151,234 & 3.7\% & 57.4\% \\
r/techsupport & 100,052 & 2.5\% & 11.1\% \\
r/instagram & 88,287 & 2.2\% & 36.9\% \\
r/legaladvice & 66,588 & 1.6\% & 7.5\% \\
r/advice & 63,871 & 1.6\% & 3.8\% \\
r/facebook & 57,695 & 1.4\% & 49.6\% \\
r/help & 47,785 & 1.2\% & 25.9\% \\
r/privacy & 41,283 & 1.0\% & 58.9\% \\
r/coinbase & 40,884 & 1.0\% & 39.2\% \\
r/nostupidquestions & 40,407 & 1.0\% & 2.3\% \\
\bottomrule
\end{tabularx}

\label{tab:top-10-volume}
\end{table}

\paragraph{Specialized help-seeking subreddits} Table~\ref{tab:top-10-density} shows the top 10 subreddits ranked by help-seeking density. Each of these communities narrowly focused on specialized support for a specific harmful experience (\eg \subreddit{r/sextortion}, \subreddit{r/computerviruses}), tool (\eg \subreddit{r/yubikey}, \subreddit{r/adguard}), or best practice (\eg \subreddit{r/cybersecurity\_help}). Consequently, they were much smaller by volume and cumulatively contributed just 3\% of help-seeking posts overall. For example, \subreddit{r/sextortion} had only 42K members and was tailored to ``current or former victim[s] of sextortion or online blackmail''. Users of the subreddit asked about the sophistication of scams (\postquote{I asked her to send me a video of her saying my name, which she did! Do scams sometimes involve a real girl?}) and how to navigate the demands of criminals (\postquote{He’s been threatening to send the photos to my school twitter if I don't send him money}). Likewise, \subreddit{r/yubikey} had 43K members, with questions touching on capabilities (\postquote{Can you use a Yubikey to unlock a phone?}) and risks (\postquote{Is it safe to use my Yubikey on a public and/or compromised computer?}). These specialized support communities likely provide an outlet to users whose questions may go beyond the expertise of general support subreddits.

Beyond the top 10 specialized communities, only 75 subreddits in our dataset had a help-seeking density greater than 50\%. Just 205 had a density greater than 25\%. These 205 subreddits cumulatively contributed only 30\% of all help-seeking posts. The remaining 70\% of posts were dispersed across 5,225 low-density subreddits---evidence that users might struggle to identify the right community to reach out to with their questions and that these communities likely do not specialize in digital privacy, safety, or security support. This poses a potential community design challenge for matching users to the necessary expertise required to answer their question. It also means that research studies for help seeking that limit their search to specialized communities may fail to discover the bulk of pertinent help-seeking needs.

\begin{table}[t]
\centering
\caption{The top 10 subreddits according to help-seeking \emph{density}, limited to subreddits that contribute at least 0.1\% of overall help-seeking posts. We additionally report the overall volume of help-seeking posts per subreddit and the overall percentage of help-seeking posts the subreddit contributes.}
\footnotesize

\begin{tabularx}{0.8\columnwidth}{Xrrr}
\toprule
\bf Subreddit & \bf{Volume} & \bf{Percentage} & \bf{Density} \\
\midrule
r/veracrypt & 3,082 & 0.1\% & 89.5\% \\
r/antivirus & 35,317 & 0.9\% & 87.7\% \\
r/computerviruses & 8,956 & 0.2\% & 84.7\% \\
r/yubikey & 5,031 & 0.1\% & 82.4\% \\
r/cybersecurity\_help & 19,280 & 0.5\% & 82.2\% \\
r/malwarebytes & 3,523 & 0.1\% & 80.5\% \\
r/identitytheft & 4,423 & 0.1\% & 80.1\% \\
r/adguard & 7,966 & 0.2\% & 76.7\% \\
r/tails & 5,384 & 0.1\% & 74.8\% \\
r/sextortion & 30,349 & 0.7\% & 74.0\% \\
\bottomrule
\end{tabularx}

\label{tab:top-10-density}
\end{table}

\begin{figure*}[t]
    \centering
    \includegraphics[width=1.0\textwidth]{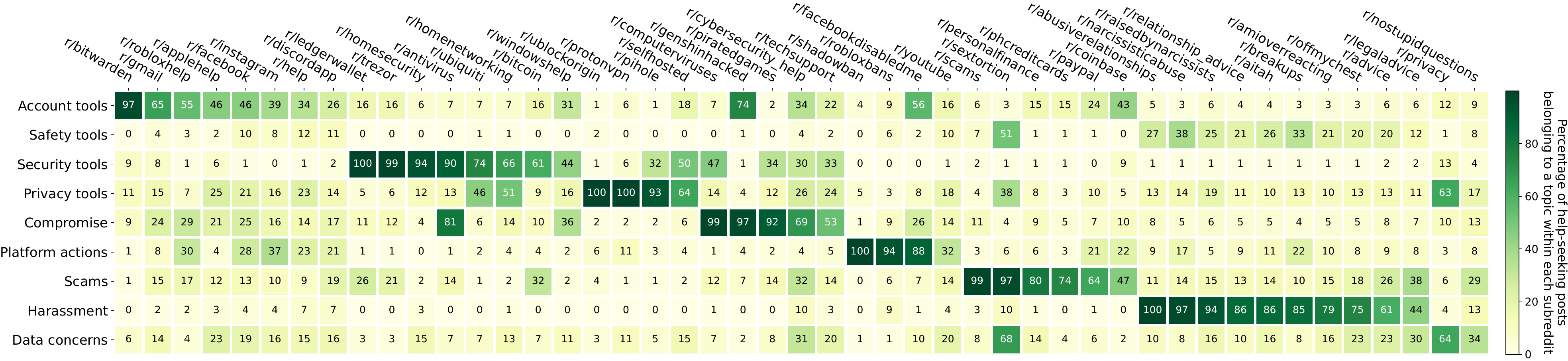}
    \caption{Percentage of help-seeking posts belonging to a topic within each subreddit. Since a Reddit post can discuss multiple help-seeking topics, the percentages in a column can sum to $>$100\%.}
    \label{fig:subreddit_heatmap}
    \Description{A heatmap showing the topics covered by different subreddits.}
\end{figure*}

\begin{figure*}[t]
\centering

\begin{subfigure}[b]{0.33\textwidth}
         \centering
         \includegraphics[width=1.0\columnwidth]{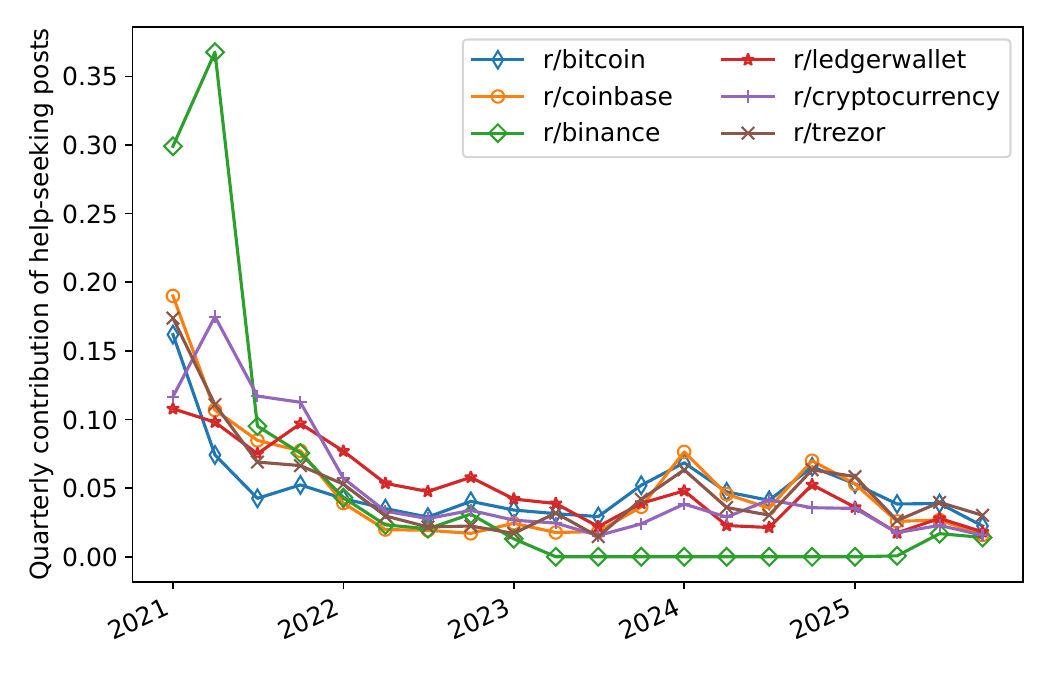}
         \caption{Cryptocurrency \& wallets}
         \label{fig:crypto_subreddit_trends}
\end{subfigure}
\begin{subfigure}[b]{0.33\textwidth}
         \centering
         \includegraphics[width=1.0\columnwidth]{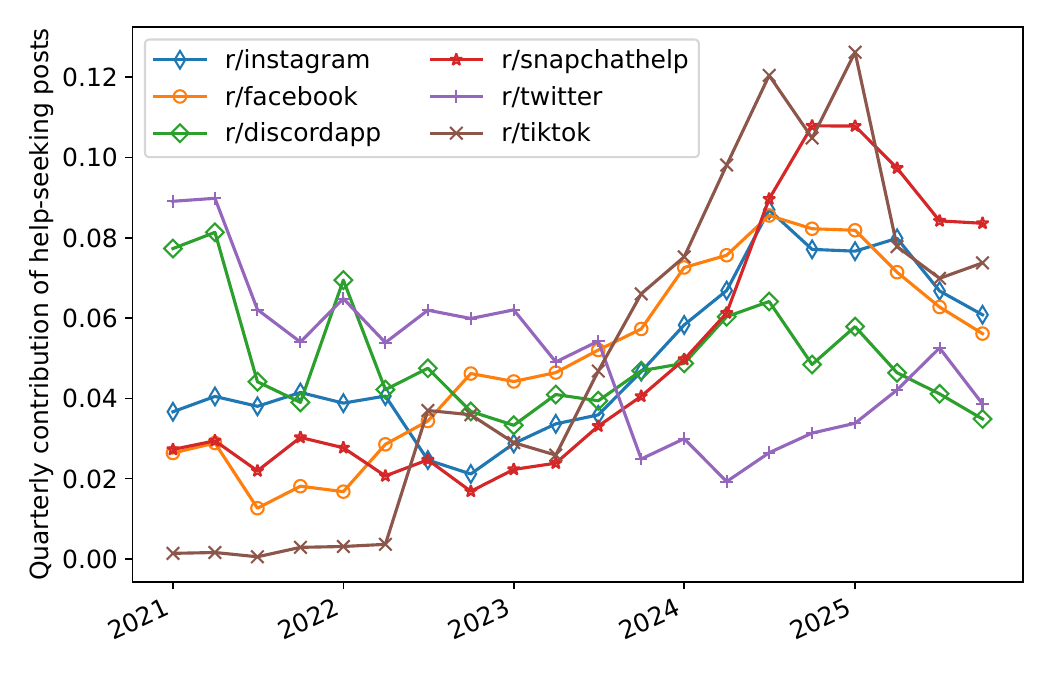}
         \caption{Social media}
         \label{fig:social_media_subreddit_trends}
\end{subfigure}
\begin{subfigure}[b]{0.33\textwidth}
         \centering
         \includegraphics[width=1.0\columnwidth]{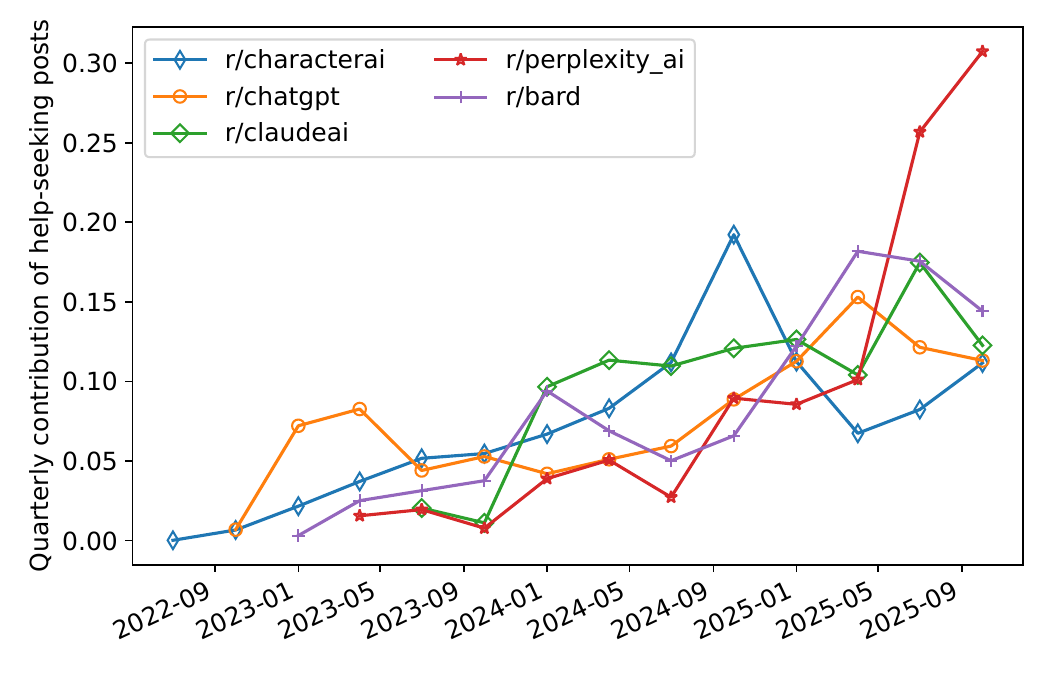}
         \caption{Artificial intelligence}
         \label{fig:ai_subreddit_trends}
\end{subfigure}

\begin{subfigure}[b]{0.33\textwidth}
         \centering
         \includegraphics[width=1.0\columnwidth]{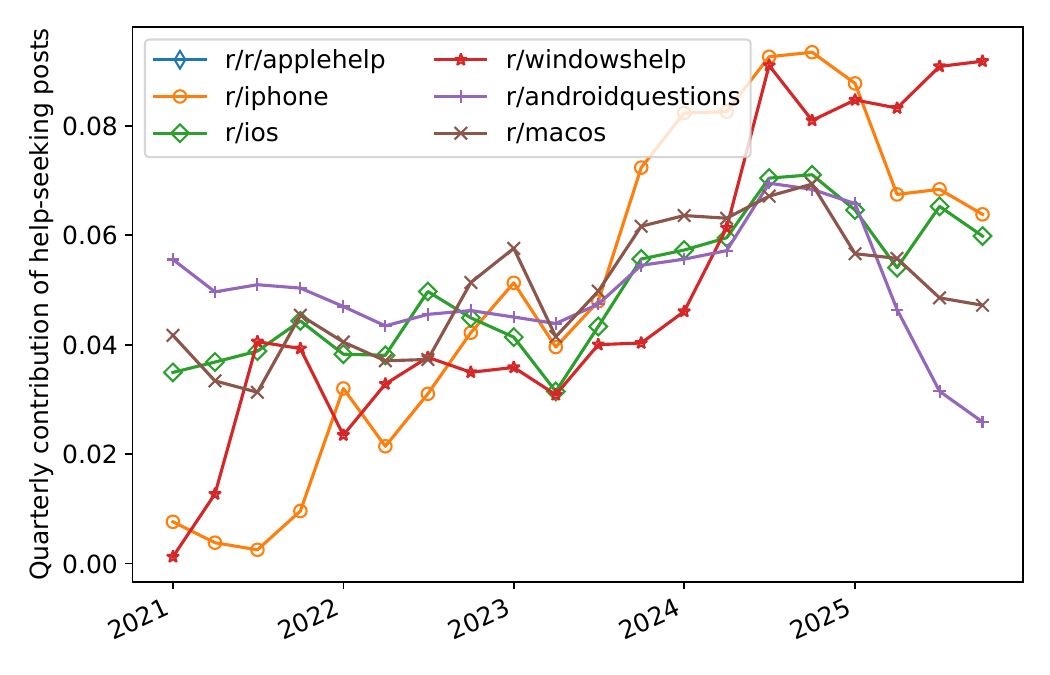}
         \caption{Device \& OS platforms}
         \label{fig:platforms_subreddit_trends}
\end{subfigure}
\begin{subfigure}[b]{0.33\textwidth}
         \centering
         \includegraphics[width=1.0\columnwidth]{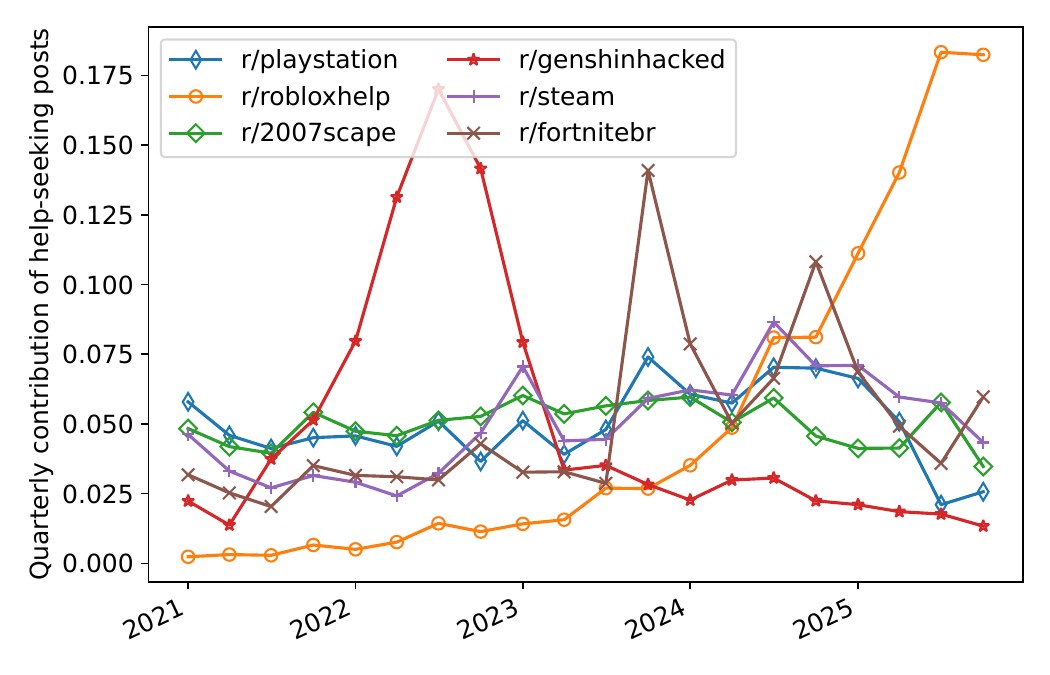}
         \caption{Gaming}
         \label{fig:gaming_subreddit_trends}
\end{subfigure}
\begin{subfigure}[b]{0.33\textwidth}
         \centering
         \includegraphics[width=1.0\columnwidth]{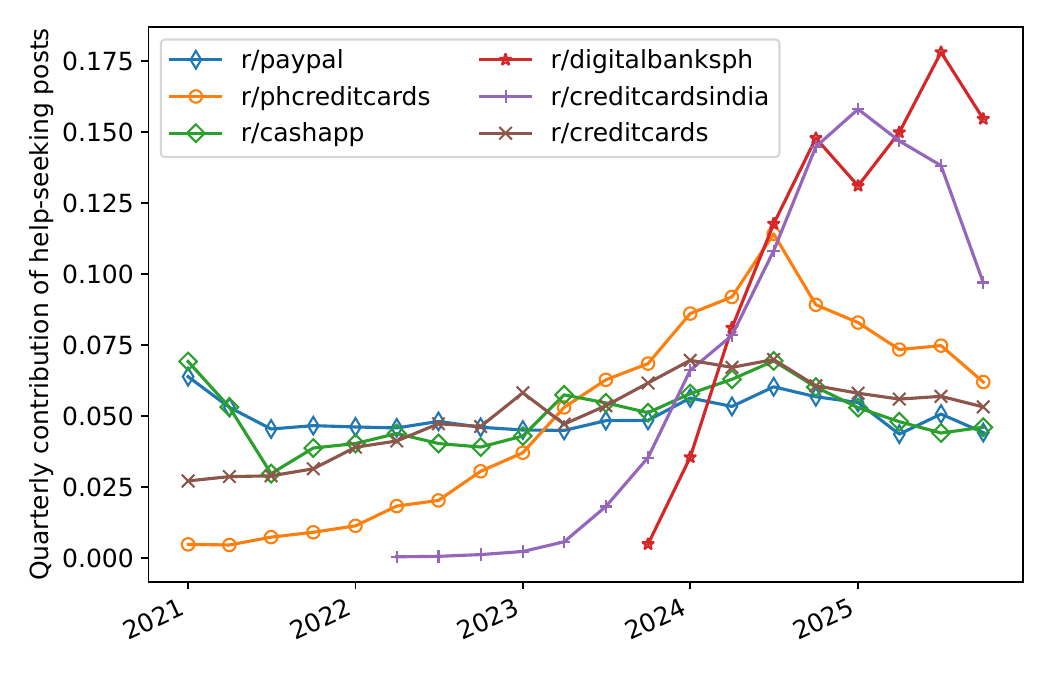}
         \caption{Payment tools}
         \label{fig:finance_subreddit_trends}
\end{subfigure}

\caption{Growth and decline of help seeking on subreddits for various technologies. Every line reflects the relative volume of help-seeking posts per quarter, per subreddit.}
\label{fig:subreddit_trends}
\Description{Six temporal plots showing the growth and decline of different subreddits.}
\end{figure*}

\paragraph{Breakdown of subreddits by help-seeking topic} We evaluate how help-seeking topics vary by subreddit in Figure~\ref{fig:subreddit_heatmap}. For each topic we extracted the top 10 subreddits by volume, which after deduplication yielded 47 subreddits (x-axis). Each cell represents the fraction of help-seeking posts discussed per topic on the y-axis. We order subreddits by the most popular topic per subreddit.

The aforementioned dynamics of general support, platform support, and specialized support are apparent in this visualization. For example, general support communities like \subreddit{r/nostupidquestions} and \subreddit{r/help} receive questions covering all nine topics. Some users appear to self-select which general support community they engage for help: Harassment was popular on \subreddit{r/relationship\_advice} (86\%); whereas Scams was popular on \subreddit{r/personalfinance} (80\%). This may stem from a perceived match between a community's (non-digital) expertise and the remedy sought by the user (\eg relational, financial).

Topics on platform support communities matched the capabilities of the platform. For example, help-seeking posts on \subreddit{r/facebook} primarily asked about Account tools (46\%), but extended to Platform actions (28\%), Compromise (25\%), and Privacy Tools (21\%). For \subreddit{r/coinbase}, questions related to Scams (47\%), Account tools (43\%), and Platform Actions (22\%). Similar trends can be observed for \subreddit{r/applehelp}, \subreddit{r/discordapp}, and \subreddit{r/gmail} among others.

Specialized support communities covered much narrower ranges of topics. We find that posts to \subreddit{r/antivirus} focused primarily on Security tools (90\%) and Compromise (81\%); posts to \subreddit{r/scams} focused on Scams (99\%); and posts to \subreddit{r/protonvpn} focused on Privacy tools (100\%). As a whole, our measurements suggest that a user seeking help has to discover which of multiple relevant subreddits best meets their needs. Each subreddit also needs members to triage multiple support topics.

\paragraph{Temporal technology trends}
Our five years of data allows us to examine how help-seeking for various technologies has evolved through time. For our analysis, we examined the top 100 subreddits by volume of help-seeking posts and manually clustered them into technology segments (\eg social media, cryptocurrency, gaming).\footnote{For AI technologies, we needed to search beyond the top 100 subreddits, as the communities have yet to grow in volume compared to other technologies.} We then measured the growth and decline of help-seeking posts over time per subreddit, calculated as the number of posts per quarter divided by the total number of posts to the subreddit in our measurement window. Figure~\ref{fig:subreddit_trends} shows our results for a selection of technology segments. We find help seeking on cryptocurrency-focused subreddits peaked in 2021 (Figure~\ref{fig:crypto_subreddit_trends}), aligning with a period when BTC rose 400\% in value. Conversely, help seeking on AI-focused subreddits continues to grow (Figure~\ref{fig:ai_subreddit_trends}) alongside rapid adoption. We also observe steady growth in help-seeking posts on \textit{r/tiktok} and \textit{r/snapchathelp}, coupled with a decline in posts on \textit{r/twitter} (Figure~\ref{fig:social_media_subreddit_trends}). Other trends include a burst in help requests related to hacked gaming accounts on \textit{r/genshinhacked} in 2022 (Figure~\ref{fig:gaming_subreddit_trends}) fueled by credential-stealing malware~\cite{kan2022genshinanticheat}, and increasing growth in payment-related questions from a variety of regions including the Philippines and India (Figure~\ref{fig:finance_subreddit_trends}) tracking mobile-wallet booms like UPI and GCash. These trends underscore how help seeking tracks the broader technology ecosystem.

\section{Help-seeking needs}
\label{sec:qual_analysis}

We conclude our results with a qualitative deep dive into help-seeking needs, with the goal of informing effective responses or solutions that meet these user needs. To that end, we examine the type of support requested (\eg factual details, guidance on next steps, reassurance), the domain knowledge necessary to interpret users' questions, and other contexts within help-seeking posts that shape users' needs.

\subsection{Types of help sought}
Our qualitative analysis of 250 posts (Section~\ref{sec:method-needs}) surfaced three primary reasons that users sought \pss help: for sensemaking, guidance, or therapeutic support. Multiple needs might appear in a single post. We discuss each of these below. While some users sought to connect with external support (\eg from an employee at a platform or law enforcement), we omit this from discussion due to its low prevalence (just 9 of 250 posts or 4\%).

\paragraph{Sensemaking} Users most often sought help in the form of sensemaking to better understand the specifics of a \pss topic (164 of 250 posts or 66\%). Example questions were akin to \postquote{what is happening?}, \postquote{anyone else ever see this?}, \postquote{anyone know why this happened?}, or \postquote{is this safe?}. These questions were often paired with context related to an issue or threat a user was experiencing. For example, a user on \subreddit{r/translationstudies} asked whether a job listing for a translator was trustworthy given the sensitive details requested:

\blockpostquote{Trying to find work\dots but scared one of [the job listings] is a scam\dots  They replied asking to do an online interview\dots 
They sent me a list of questions to answer\dots [Lists questions]\dots I don't really understand why they might need to know which phone carrier I use or if I have a good credit score or not\dots Sorry if I'm just being paranoid, but I'm just trying to be cautious while finding work online!”}

{\noindent}Apart from threats, users would also seek help understanding technology issues. For example, a user on \subreddit{r/bitwarden} asked:

\blockpostquote{I created a Bitwarden account, downloaded Firefox extension and Android app and logged in on both. Then I added 2FA via passcode, for which I use Duo. That worked on Firefox, but then I couldn't sync on Android.\dots What's going on?}

Other sensemaking needs expressed by users included requests for factual information, explanations, recommendations, predictions, and subjective judgments of situations or technology. Addressing sensemaking needs requires synthesizing a user's unstructured concerns and providing explanations and rationales.

\paragraph{Guidance}
\label{sec:needs-guidance}
Users also sought guidance on actions they could follow to resolve \pss issues (100 of 250 posts or 40\%). These requests were often paired with descriptions of actions already taken by the user or detailed context about the issue. Some of these posts touched on specific procedures where there was likely a knowable, correct answer. For example, one post to \subreddit{r/techsupport} asked about how to configure system updates (which would include security updates) after jailbreaking their device:

\blockpostquote{I just realized my Amazon tablet hasn't updated since I got it in 2020. `Check for updates' function in settings just fails instantly. This is probably a result of me `jailbreaking' it to force install  Google Play Store. I used the Amazon Fire Toolbox V9.2.1 to do this. However the toolbox doesn't have an option to update the Android OS. How do I go about this?}

However, many guidance questions did not have straightforward solutions. For example, one user wanted help contacting a platform for support:

\blockpostquote{Is there any actual way to contact this game support? My account has been stolen and the stupid account retrieval form is literally useless and when i try to contact them via email it auto replies that this email is only for login stuck issues and for other issues use game support.}

Addressing guidance needs likely requires taking into consideration a user's context, asking follow-up questions, and helping the user move forward. It also requires recognizing when there is no viable solution.

\paragraph{Therapeutic} 
While less common, users also sought therapeutic support (47 of 250 posts or 19\%). Users authoring these posts were seeking emotional reassurance (\postquote{Please help put my mind at ease}), an outlet for strong emotions or to tell their story (\postquote{I feel really hopeless}), commiseration (\postquote{I hate this site}), or validation that their emotions were a reasonable response to a frustrating or painful situation (\postquote{I'm really just looking for support or to vent}). Most of these users were also seeking sensemaking or guidance help. Some users recognized that their situations were incredibly challenging and were not necessarily seeking concrete guidance to resolve the situation (though they might include a plea for anything that could improve it).  

Addressing therapeutic needs goes beyond technical reassurance, recommendations, or step-by-step guides and likely requires a sense of community. Even if a help resource does not directly address therapeutic needs, it likely needs to provide sensitive, empathetic support. 

\subsection{Critical domain knowledge}
Help-seeking posts commonly required domain-specific knowledge. For example, a majority of posts touched on specific platforms and tools as part of the scope of help seeking (142 of 250 posts or 57\%). Users referenced multi-purpose platforms (e.g., Google, Apple, Microsoft), messaging and social media (e.g., Instagram, Facebook, TikTok, WhatsApp, Snapchat, Reddit), crypto platforms or wallets (Blockfi, KuCoin, Ledger Nano, Binance), peer-to-peer money apps (e.g., CashApp, Venmo, PayPal), games (e.g., Roblox, Fortnite, RuneScape, Playstation), security and privacy tools (e.g., Wireguard, Windscribe, Bitwarden, Opacity), shopping apps,  (e.g., Temu, eBay, Etsy), and many more. As an example, some users sought comparisons between the \pss merits of different tools or platforms, 
such as this post to \subreddit{r/electrum} on whether to use a hardware or software cryptocurrency wallet:

\blockpostquote{Want to make sure I understand the pros and cons of these two approaches: Use BitBox app alone (without Electrum, using vendor software) to generate wallet. Use Electrum to generate wallet and store seed on BitBox. Is one inherently safer / better than the other?}

Knowledge of attackers, digital-safety risks, and tech-facilitated harms was also commonly required to interpret help seeking requests (111 of 250 posts or 44\%). This stems from many questions relating to harmful experiences such as Compromise, Platform actions, Scams, Harassment, or Data concerns (Section~\ref{sec:topic_trends}). Users requested help to understand and/or determine how to act in specific threatening or harmful circumstances. For example, one user asked: 

\blockpostquote{Does anyone know why hackers are interested in my Spotify account? I don't have Spotify premium, never have. Literally nothing valuable on my ACC\dots and somebody suspicious has logged in TWICE?}

Other pertinent domain knowledge related to technical needs (39 of 250 posts or 16\%) as well as non-technical needs including financial advice (37 or 15\%), relationship advice (16 or 6\%), and legal advice (8 or 3\%). The sheer breadth of domain knowledge underscores the complexity of connecting users with the right expertise for their question.

\subsection{Contextual details} 
Help-seeking posts often referenced highly-specific contextual details that users interlaced with their questions. Two important contexts that we observed were \emph{root causes} and \emph{practices}. Root causes referenced experienced, threatened, or otherwise non-hypothetical issues that precipitated the help-seeking post. Illustrative examples from different posts included \postquote{my [Pixel] 5a hasn't been receiving any updates since Aug 5} and \postquote{noticed ES File Explorer taking up nearly 16gb of storage} after which this user suspected it was stealing their data. Separately, practices reflected a user's existing \pss posture or steps they previously took to remediate an issue. Examples included \postquote{I try to recover it, it asks me what my last password was}; \postquote{I blocked a contact on my phone}; and \postquote{anti-virus says all the APK I downloaded were safe}. 

Providing tailored help can thus require interpreting lengthy contexts, ignoring tangential details, identifying potential inaccuracies or omissions to probe deeper on, and ultimately recommending a course of action or other form of support. In the following example (demonstrating Compromise as a root cause and Account Tools use as practices), generic account security advice would be unsatisfactory to this user who has exhausted existing options (and who may be infected with malware):

\blockpostquote{I’ve had the unfortunate luck of having my jagex [gaming platform] account hacked [\textbf{root cause}]. I’ve changed the password  [\textbf{practice}] and kicked all the accounts off the launcher requiring a re log  [\textbf{practice}]. I have set up a 2fa [\textbf{practice}] and feel the account is safer now however they’re still able to log in and be on the account [\textbf{root cause}] which I’m unsure how that’s possible and wondering if anyone could help or give tips.}

\section{Discussion}
Our results illustrate the complexity of \pssand help seeking across Reddit. We discuss ways that help-giving resources can potentially overcome at least some of this complexity.

\subsection{Complex help-seeking needs} Our measurements demonstrate that help seeking reflects
a complex combination of threats, platforms, mitigations, context, and emotions. Users' needs are co-evolving alongside attackers' tactics and platform capabilities. Users' help-seeking questions are not effectively answered by simply enumerating the benefits of two-factor authentication or outlining the tradeoffs of various privacy settings. Instead, help seeking is richly contextual, capturing the state of a user's comprehension of a topic, their suspicions for the root cause of an issue, and detailed technical descriptions of their system environment and prior remediation attempts. Addressing these needs requires synthesizing the user's unique context; having intimate knowledge of platform idiosyncrasies; having expertise across digital privacy, safety, and security risks and harms; and iteratively providing instructions---all while users experience confusion, annoyance, and even fear that may stymie their forward progress. 

\subsection{Design implications}
Our measurements show that \pss help seeking on Reddit includes millions of bespoke questions. We posit this is likely due to inadequate existing self-help resources, because they might be undiscoverable, unwieldy, or unhelpful. Potential solutions include building novel help-giving agents or improving existing help-giving pathways. Our implications are organized for three key stakeholders in help-giving: LLM developers, security and privacy advice guide writers, and community organizers, including Reddit mods.

\paragraph{LLM agent developers}
Agents---LLMs that use tools to engage in multi-step reasoning to solve tasks---represent a promising design direction for help giving. Agents appear to be well-suited to the rich contextual details present in our dataset. They also provide an experience similar to Reddit that allows for natural language queries and interactive replies. Multi-modality might allow an agent to interpret screen contents (\eg current permission settings), or even configure those settings with app or browser integration. Agents could potentially collaborate with and integrate expert-written guides (discussed below), which may help to reduce the risk of an LLM offering incorrect advice. Similarly, agents might benefit from platform-specific knowledge, something that prior research indicates current frontier models struggle with~\cite{prakash2024assessment}. However, agents may be unsuitable to meeting the therapeutic needs of users---such as their desire for emotional support and a sense of community. 

Properly evaluating how LLMs respond to the nuanced help-seeking questions we have extracted constitutes a significant undertaking that is outside the scope of this paper. Constructing a rigorous evaluation framework requires developing comprehensive rubrics capable of capturing the multifaceted dimensions of an LLM's reply. Such rubrics must assess not only technical accuracy and readability, but also evaluate an agent's ability to appropriately manage alarming commentary, accurately gauge a user's level of technical expertise, explain risk tradeoffs, determine exactly when it is necessary to route a user to a dedicated human or help center, and more. Our dataset and insights detailing the context of how users seek digital privacy, safety and security help could be leveraged to evaluate
 benchmarks---which are actively being developed~\cite{prakash2024assessment, tihanyi2024cybermetric, liu2024cyberbench,jing2024secbench, meiklejohn2026helpbench}---to reflect the real-world complexity of help seeking.

\paragraph{Guide writers}
Our results indicate that help seeking tends to narrowly focus on just one topic at a time (\eg Scams vs. Privacy tools).\footnote{Over half of the posts we found (58\%) exclusively focused on one topic; 91\% contained one or two topics.} This allows online resources and help centers to potentially specialize in which topic they support. As always, this comes with the caveat for whether experts for a given topic agree on best practices. While account security advice is prescriptive among experts~\cite{securityadvice2017}, emerging threats like harassment lack the same consensus around protective practices~\cite{wei2023there}. Similarly, while privacy tools and data concerns were among the most popular help-seeking topics in our study, it may be difficult for a single privacy guide to capture every user norm~\cite{dupree2016privacy}. 

Beyond topics, guide writers should also consider the substance of the help they offer. Some guidance needs (\eg requests for step-by-step instructions) are well-suited for online resources or help centers. But users also had more complex guidance needs---like asking for advice in situations where there is no agreed upon approach---and sensemaking needs---a desire to know what is going on and what to make of their situation. Therapeutic needs may be ill-suited for such resources. Writers should also consider tailoring resources per platform. Navigating this level of specificity (\eg topics and platforms) may require restructuring how users search for help, with users providing a description of their issue which triggers a semantic-based search, rather than keyword-based searches or topic indexes.

\paragraph{Community organizers}
Help giving on Reddit is fragmented. Users have to identify which of thousands of subreddits best meets their bespoke questions. Meanwhile, every subreddit needs to have community members with the expertise and bandwidth to answer tens of thousands of questions, or risk having a smaller set of experts spend considerable time correcting others or having questions go unanswered. These issues of discovery and scaling are not unique to Reddit and impact other offline service providers that help people to navigate \pssand (\eg advocacy groups, social workers)~\cite{havron2019clinical, tseng2022care}. In the context of Reddit, automation may help with the intake of users seeking help. For example, \subreddit{r/scams} provides community members with macros to automatically provide a user seeking help with descriptions of popular attacks (\eg airdrop scam, romance scam, tech support scam). 

Further research is needed for how best to blend all the aforementioned resources to improve the quality of help, creating an ecosystem of care that can meet all of the sensemaking, guidance, and therapeutic needs of users, without overburdening any single pillar of support.

\section{Conclusion}
We presented the results of a mixed methods study of five years of help seeking for \pssand across Reddit. We built an analysis pipeline that sifted through two billion Reddit posts from 2021--2025 to identify posts where users asked for \pss help. We identified four million relevant help-seeking posts with 93\% precision and recall. We automatically annotated these with the topics discussed (\eg security tools, privacy configuration, scams, and more) and the primary emotional sentiment of the post. Analyzing this help-seeking dataset, we found that yearly volume of help seeking grew by 52\% in the last 5 years, reaching a peak of 102,000 questions per month at the end of 2024. Users asked for help for a variety of topics, some of the most popular of which included scams (29\%), account tools and access issues (21\%), and privacy tools and settings (19\%). Our qualitative analysis showed that providing help seekers with effective support requires navigating a complex combination of threats, platforms, mitigations, contexts, and emotions. We discussed how LLM agents, guide writers, and community organizers could operate collaboratively as an ecosystem of care to better meet all of a user's \pssand help-seeking needs. To support such advancements, we have open sourced our help-seeking dataset.\footnote{https://github.com/google/helpseeking}

\appendix

\begin{acks}
This research received no specific grant from any funding agency in the public, commercial, or not-for-profit sectors.
\end{acks}

\begin{ethics}
Our work does not directly recruit or engage with participants. Our study plan was reviewed by experts at our institution in domains including ethics, human subjects research, policy, legal, and privacy. While we do not have an IRB, we adhere to similarly strict standards.

\subsubsection*{Purpose \& decision}
We approached this research project with a focus on improving help-giving resources for digital privacy, safety, and security---a benefit for billions of technology users globally. Such improvements require collecting data to understand the sources of friction that technology users experience. Reddit provides one such lens with a broad user base. Furthermore, rapid changes in technology and risks necessitates collecting this data over a longitudinal period to ensure any insights or design recommendations are stable and thus might provide lasting utility. Benefits from our research include (1) improving the research community's, platforms', and regulators' understanding of digital privacy, safety, and security help needs; (2) the introduction of new methods that scale qualitative methods in order to study massive datasets; and (3) the public release of millions of help-seeking posts to support future research and development surrounding help-giving resources.

\subsubsection*{Stakeholders \& risks}

\paragraph{Researchers} Researchers engaging in this study were exposed to sensitive topics pertaining to digital privacy, safety, and security---a potential risk to their well-being. These researchers also have a vested interest in sharing findings and improving digital security, privacy, and safety.

\paragraph{Reddit users} Users posting publicly to Reddit may be unaware that the content they share---such as personally identifying \newline information---is used for research.

\paragraph{Reddit platform} Reddit's Terms of Service lay out strict requirements of how content on its servers can be accessed or shared.

\paragraph{Broader technology platforms} Reddit users may mention technology platforms tied to negative privacy, safety, or security \newline experiences---a potential risk to the platform's reputation. Platforms may also benefit from insights into how to improve their offerings.

\paragraph{Governments and regulators} Governments and regulators may have a vested interest in the digital security, privacy, and safety of technology platforms. 

\paragraph{Research community} The broader research community has a vested interest in improving digital security, privacy, and safety, and that any shared datasets are sound.

\paragraph{Society} Society writ large can benefit from improvements to digital security, privacy, and safety.

\subsubsection*{Mitigations}

\paragraph{Restricting analysis to public data} Our study focuses on publicly accessible data on Reddit---with ethical guidelines on how to handle such data covered in greater depth by Fiesler \etal~\cite{fiesler2024remember}. We rely on a crawler to access all data in our study in a way that respects \texttt{robots.txt}. As part of our effort to mitigate any harm to Reddit users, we excluded posts in our original crawl belonging to communities that were subsequently made private, quarantined, or banned at the time of our analysis.

\paragraph{Data minimization} When quoting posts, we omit reference to the user that authored the post and redact personally identifying information shared by the user. We retain the names of platforms that appear in quotes and subreddit titles as they are crucial to understanding the nuance surrounding a user's experience. However, we avoid selecting quotes in our paper that are explicitly critical of any platform, in order to minimize any perceived bias in how the research team might have selected such quotes, and to minimize any reputational harm to platforms.

\paragraph{Data re-hydration} To support future research into help-seeking, we make available our golden datasets all help-seeking posts identified by our pipeline, along with label annotations (see Open Science for more details). However, Reddit Terms of Service preclude sharing the content of these posts, and users may delete the content of their post in the future. To minimize harm to both of these stakeholders, we adopt a re-hydration strategy where we provide only the URL and all the annotations generated by our pipeline. Other researchers will thus have to download the content at the URL in a way that respects Reddit's controls.

\paragraph{Automated labeling} In order to reduce the toll on researcher well-being when examining potentially sensitive user experiences tied to digital privacy, safety, and security, we relied heavily on automation through LLMs to scale qualitative research methods to the hundreds of millions of posts in our dataset. While our study focuses on 4 million posts, researchers involved in the effort reviewed roughly 3,000 posts through stratified sampling.

\paragraph{Well-being practices} We followed similar well-being practices to those documented by Wei \etal~\cite{wei2024understanding} such as group check-ins, meeting with trauma-informed experts, having access to therapists, taking breaks, and (as just described) reduction of exposure through automation. Bellini \etal offer a richer treatment of best practices when engaging in similar research~\cite{bellini2024sok}.

\end{ethics}

\begin{openscience}

\paragraph{Reddit crawl} Access to Reddit data is governed by its policies, which precludes publicly redistributing the content of posts and other associated metadata. As such, we are unable to make available to the community the two billion posts we crawled. Due to technology restrictions, we are also unable to share the crawler used in our study. Researchers seeking to replicate this infrastructure may consider alternatives such as Reddit's API or Pushshift, though need to familiarize themselves with the data restrictions of those tools.

\paragraph{Codebooks \& golden datasets} Our definition for help seeking was previously laid out in Section~\ref{sec:methods}. We make our codebook for differentiating the help-seeking topics available in our Appendix. Researchers in our study used this codebook to determine which of nine possible help-seeking topics were present in a post. We make our golden dataset for our help-seeking definition and for our help-seeking topics available at \url{https://github.com/google/helpseeking}. As discussed in our Ethics Considerations, these datasets do not contain any content; we only provide permalinks to access the content.

\paragraph{Prompts \& models} We make all prompts used for identifying help-seeking posts and annotating help-seeking topics available in our Appendix. We used these prompts to fine-tune a Gemini 1.5 and Gemini 2.0 Flash model. Due to restrictions on Gemini as a closed-source model, we are unable to share the model weights or LoRA weights. Researchers seeking to replicate this infrastructure can use Google Vertex AI's tuning API,\footnote{https://docs.cloud.google.com/vertex-ai/generative-ai/docs/models/tune-models} other closed-weight models with tuning interfaces, or open-weight models. We previously laid out our model tuning parameters and training, validation, and testing breakdowns in Section~\ref{sec:methods}.

\paragraph{Annotated Datasets}
Our final annotated dataset is available at \url{https://github.com/google/helpseeking}. As discussed in our Ethics Considerations, this dataset does not contain any content; we only provide permalinks to access the content.

\end{openscience}

\begin{ai}
The authors used generative AI-based tools to identify misspellings and grammatical errors. As described in Section~\ref{sec:methods}, we used AI for automatically identifying and annotating relevant help-seeking posts. We have manually verified and are responsible for the accuracy,
originality, and integrity of the output of all AI-based tools.
\end{ai}

\bibliographystyle{ACM-Reference-Format}
\bibliography{references}

\section{Golden dataset sampling}
\label{sec:golden_dataset_sampling}
\begin{figure}[t]
\centering

\begin{minipage}{\columnwidth}

   \begin{tcolorbox}[colback=green!5!white,
                    colframe=green!20!white,
                    arc=4mm,
                    ]
     \footnotesize {You are an expert assistant designed to identify people discussing or asking questions related to a specific topic. Please review the following Reddit post and answer Yes or No if the author is discussing or asking any question related to:\\
     
- End-to-end encryption\\
- Private messaging\\
- HTTPS\\
- VPN\\
- Tor\\
- Anonymity\\
- Incognito mode\\

{\bf Subreddit:} [SUBREDDIT]\\
{\bf Reddit post:} [POST]\\
{\bf Answer:}
}
  \end{tcolorbox}
\end{minipage}
\caption{Stratified sampling prompt designed to identify posts related to a specific privacy, safety, or security topics---in this case, encryption or anonymity tools.}
\label{fig:prompt_stratified_sample}
\Description{The prompt text for identifying security, privacy, and safety related posts.}
\end{figure}

\paragraph{Help-seeking subreddit sampling (N=800)} We randomly sampled 800 posts across the following subreddits: 
\seqsplit{r/abusiverelationships, r/advice, r/ask, r/askgaybros, r/askmen, r/asknetsec, r/askreddit, r/asktransgender, r/askuk, r/care, r/cptsd, r/cyberlaws, r/cybersecurity, r/cybersecurity101, r/cybersecurity\_help, r/dating, r/dating\_advice, r/depression, r/domesticviolence, r/eff, r/emotionalabuse, r/europeprivacy, r/facebook, r/gdpr, r/hacking, r/help, r/howtohack, r/infosec, r/legaladvice, r/makemefeelbetter, r/mentalhealth, r/narcissisticabuse, r/opsec, r/privacy, r/privacyguides, r/privacytools, r/privacytoolsio, r/privatelife, r/raisedbynarcissists, r/redditsecurity, r/relationship\_advice, r/relationships, r/scams, r/security, r/sex, r/sextortion, r/sexualassault, r/stalking, r/suicidewatch, r/technologyprotips, r/techsupport, r/trueoffmychest, r/twoxchromosomes.}

\paragraph{Privacy, safety, and security topic sampling (N=281)} We used Gemini 1.5 Flash to identify posts discussing privacy, safety, and security, irrespective of whether a user was seeking help for the topic discussed. We constructed 14 distinct prompts covering 79 different topics (\eg end-to-end encrypted messaging, account takeover, permission settings, \etc). We show one prompt variant in Figure~\ref{fig:prompt_stratified_sample}. 
We applied each prompt to 50,000 posts selected randomly across Reddit. Of 700,000 results output by Gemini, just 0.5\% indicated the positive presence of a privacy, safety, or security topic. We selected up to 30 positively labeled posts per prompt variant, which after de-duplication yielded a final sample of 281 posts. 

\paragraph{Model-guided sampling (N=935)} We prompted Gemini 1.5 Flash with a preliminary help-seeking prompt (Appendix~\ref{sec:helpseeking_prompt}) that we applied to a random sample of 10M posts across Reddit, from which we selected 935 posts that Gemini labeled as potentially help seeking.

Our final sample, the aggregate of these three strategies, consisted of 2016 posts covering 749 subreddits. We removed 16 posts from consideration due to being cross-posted duplicates, or due to containing just a URL, the contents of which would be inaccessible to an LLM without tools. This left us with 2,000 sample posts.

\section{Help-seeking identification prompt}
\label{sec:helpseeking_prompt}

   \begin{tcolorbox}[breakable,colback=green!5!white,
                    colframe=green!20!white,
                    arc=4mm,
                    ]
     \footnotesize {
     The following text is from an online communication posted to the subreddit ``\{\{subreddit\}\}'':
``\{\{text\}\}'' \\

Based on the text content and the subreddit, we need to determine if the text is seeking help related to privacy, security, abuse, or a safety issue. A HELP SEEKING post includes two parts:\\

\textbf{\underline{PART I - GOAL}}:
The author is seeking to prevent, or is experiencing, or is close to someone who has experienced or perpetrated a tech-facilitated privacy, security, abuse, or safety issue. Posts about business needs or those by perpetrators trying to do harm or evade safety guards are OUT OF SCOPE and NOT HELP SEEKING.
We consider a post to potentially be HELP SEEKING if the issue mentioned in the post EXPLICITLY involves the use of technology (as a source of harm or source of solution) and EXPLICITLY requests for help.
Also, it should include one of the following personal aspects:\\
- \textit{Active/Prior Event}: Explicitly sharing details of the security, privacy, abuse, or safety issue the author or someone they are close to are facing or perpetrating.\\
- \textit{Preventative Event}: Content related to preventing potential harms by enabling or modifying security, abuse, or privacy settings. Educational queries focused on improving personal security or privacy postures are also considered preventative events.\\

\textbf{\underline{PART II - TYPE OF HELP SOUGHT}}: We consider a post to be help-seeking if it explicitly includes one or more of the following. Requests outside these issues (e.g., requests for funds) are OUT OF SCOPE and NOT HELP SEEKING.\\
- \textit{Sensemaking}: Asks or expresses a need for support in trying to understand what is happening, or in understanding an issue or related resource (a technology, setting, process, etc.).\\
- \textit{Guidance}: Asks or expresses a need for guidance about what self-reliant actions they can take next. This includes technical advice, or whether or how to use platform-provided tools or help pages.\\
- \textit{Therapeutic}: Asks or expresses a need for validation or emotional support related to an issue.\\
- Relationship: Asks or expresses a need for advice on how to navigate a relationship affected by the issue.\\
- \textit{External support}: Asks or expresses a need for advice on whether or how to reach out to a third-party for formal or informal support, including legal, platform employees, social workers, etc.\\
- \textit{Other}: Asks or expresses a need for some other type of help that pertains to a tech-facilitated privacy, security, or safety issue not covered by the types above.\\

We consider a post to be NOT HELP SEEKING for tech-facilitated privacy, security, abuse, or safety issues if it matches any of the following:\\
- A privacy, security, abuse, or safety issue but does not have a technology component (i.e., is not a tech-facilitated issue).\\
- Has an incidental mention of tech-facilitated privacy, security, abuse, or safety issue or solution, but where the help-seeking request is not related to the technology use.\\
- Implicitly requests for support to prevent or cope with a privacy, security, abuse, or safety issue. We want a help-seeking post to explicitly request for support.\\
- Purely descriptive of an experience with an issue, without any request for support.\\
- A fantasy situation or a news story.\\
- An “Ask me anything” (AMA) where people invite questions about their experiences.\\
- A moderator warning about unacceptable behavior on the forum.\\
- A general complaint about the state of the world without an expressed desire for a specific solution or help-seeking need.\\
- Seeking help about a violation of trust that does not constitute abuse. (e.g., "I found out my boyfriend is sending nude photos to his ex")\\
- Providing advice, support, or resources to someone experiencing online abuse, such as a public service announcement warning or informing others about an abusive person or online space.\\
- Queries related to enterprises/companies, rather than an individual.\\
- An information request for purely educational purpose ("I want to grow my skills in cyber security"). To include educational requests, the post needs to explicitly be in support of preventing or coping with personal privacy, security, abuse, or safety issues.\\
- By a perpetrator who wants to harm others, or wants to evade or remedy an existing privacy, security, abuse, or safety protection/safeguard.\\

You are to provide your classification of the post as a binary (0 or 1) response, where 0 indicates the post is NOT HELP SEEKING and 1 indicates it is HELP SEEKING.\\

Below are a few hard example posts and their classification:\\
\{\{examples\}\} \\

Based on the HELP SEEKING definitions and the examples shared above, what is the binary classification for the following text and subreddit?\\
- text: ``\{\{text\}\}''\\
- subreddit: ``\{\{subreddit\}\}''
}
  \end{tcolorbox}

\section{Help-seeking topic codebook}
\label{sec:topic_codebook}

Each topic is defined below. The codebook also included examples for coders to reference, like those listed in the Topic Prompt shown in Appendix \ref{sec:helpseeking_topic_prompt}.

\textbf{Account tools}: Mentions of access to or configuration of an account, device, or service. These are typically accounts held by the user. Requires an explicit mention of access or configuration issues (not implied by a report of hacking, account banning, etc.). This code may be applied on its own or together with other Topic codes that explain what happened for the user to have account access or configuration questions (e.g., content moderation, compromise, etc.).

\textbf{Safety tools}: Mentions of tools related to moderation or countering abuse. Excludes spam and screening (apply Security Tools code instead).

\textbf{Security tools}: Mentions of tools related to security practices. 

\textbf{Privacy tools}: Mentions of tools related to privacy practices. Apply Data Concerns code for requests involving explicit mentions of data privacy concerns; this Topic code can be co-applied when a specific tool is stated as related to the request.

\textbf{Compromise}: Mentions of an account or device being compromised by an (implied or explicitly noted) attacker. This code requires a (perceived, implied, or explicit) attacker. This can include cases where described behavior strongly implies a related attack, inferred by the researchers, even if the poster does not understand the situation. Also use the Account Tools code if the user explicitly mentions issues accessing an account. Also use the Data Concerns code if the user explicitly mentions concerns about data exposure due to compromise.

\textbf{Platform actions}: Mentions of a safety control or policy (such as content moderation or fraud detection), typically by a platform or third party (such as an employer), that has gone awry and/or is causing friction with the user.

\textbf{Scams}: Mentions of fraud, scams, or impersonation for the apparent purposes of a scam or fraud. Also includes transactions and missing funds or goods where the reason is unclear. Use this code when a user encounters a potentially fraudulent artifact (like a suspicious website or message), but hasn’t yet been compromised (as opposed to the Compromise code). Use the Harassment code for impersonation for the apparent purposes of harassment or abuse.

\textbf{Harassment}: Mentions of interpersonal harassment or abuse by an attacker, including sexual abuse and grooming. 

\textbf{Data concerns}: Mentions related to a user’s concerns or unease about the way their data is being used by or exposed to others (typically a platform). Use the Harassment code instead when data is exposed or tampered with by a person, where the intent seems to be to harm someone. Use the Privacy Tools code (possibly double-coding) when they are explicitly referring to a tool or practice that would increase privacy.

\section{Help-seeking topic identification prompt}
Note that the topic definitions in the prompt are different from Appendix~\ref{sec:topic_codebook} (which continued to evolve after the prompt). This does not pose an issue as we LoRA fine-tuned using only data labeled with the most up-to-date qualitative codebook. The text definitions are only a hint to the LLM, whereas the labels are what guide the model training.

\label{sec:helpseeking_topic_prompt}

   \begin{tcolorbox}[breakable,colback=green!5!white,
                    colframe=green!20!white,
                    arc=4mm,
                    ]
     \footnotesize {
You are a helpful assistant designed to detect Reddit posts discussing privacy, security, or safety topics. Answer Yes or No for whether a post discusses any of the following topical categories, even if there is just a brief
mention.

-----------------------------------------\\
Here are the possible list of topical categories and their descriptions that aid in assignment:

\textbf{Category-1: Account tools}\\
\textit{Definition}: Mentions of how to configure or recover an account, device, or  service related to access or configuration. These are typically accounts held by the user.\\
Posts related to this category may discuss any of the following:\\
- Recovering access to an account, service, or device (due to lockout, forgotten password, or a login issue) \\
- Difficulty creating, registering, or verifying an account or service \\
- Recovering access to a cryptocurrency wallet (due to a lost or inaccessible wallet) \\
- Verifying a cryptocurrency wallet \\
- Accessing the login history of an account \\
- Restore device from backup \\
- Resetting a password, changing a password, creating a strong password, or using a password manager \\
- Configuring authentication security features like two-factor authentication, SMS authentication, one-time passwords, security questions, fingerprints, biometrics, passkeys, or pin numbers \\
- Losing access to an account, device, or service \\
- Account management settings \\
The following posts are not related to this category:\\
- Discusses a password leak, having their account blocked, or having their account hacked.\\
\textbf{Category-2: Safety Tools} \\
\textit{Definition}: Mentions of tools related to moderation or countering abuse. Any  mentions of spam, or screening, should be `Security tools' category.\\
Posts related to this category may discuss any of the following:\\
- How to block a gaming or social media account, email, or phone number\\
- How to report another user on any platform (e.g., for harassment)\\
- How to request an account be taken down on any platform (e.g., for impersonation)\\
- How to use moderation tools on any platform\\
- How someone circumvented being blocked on any platform\\
The following posts are not related to this category:\\
- Discusses spam or scams\\
\textbf{Category-3: Security tools}\\
\textit{Definition}: Mentions of tools related to security practices.\\
Posts related to this category may discuss any of the following:\\
- Security tools such as anti-virus, Virus Total, safe search, Safe Browsing, firewalls, spam filters, or caller ID\\
- Security applications such as Wireshark, SELinux, SSH\\
- Security hardware like trust platform modules or trusted execution environments\\
- Cryptocurrency wallets, KYC processes, hardware wallets, ledger\\
- Security practices like encryption, HTTPS, filesystem access control, port scanning/blocking, browser safety settings, access controls, secure mail-forwarding, end-to-end encryption, and audit logs\\
- Software or security updates\\
- Physical security measures for devices (camera and microphone covers)\\
- Network security, wifi security, device security, camera security, operating system security, jailbreaking\\
- Advice on which application is more secure\\
The following posts are not related to this category:\\
- Discusses account security or passwords\\
- Discusses VPNs\\
- Discusses privacy\\
\textbf{Category-4: Privacy tools}\\
\textit{Definition}: Mentions of tools related to privacy practices.\\
Posts related to this category may discuss any of the following:\\
- Privacy tools (e.g., Private browsing; Private operating systems; Private search engines; Tor; VPN; Ad blockers; Brave; Tails; SSH; Google Voice)\\
- Privacy practices, configuration, and hardening (e.g., Creating a secondary account; Creating a secondary phone number; Changing addresses; Anonymous accounts; Disabling syncing; Going off grid; Disposal of devices; Cookies)\\
- Privacy protocols such as encryption, E2EE, SSH, HTTPS\\
- Privacy settings and data controls (e.g., Cookies; Permissions; Settings; Deletion of data or accounts; Isolation; Data hiding) \\
- Privacy techniques like deleting data, hiding data, isolating data, deleting accounts, or preventing tracking\\
- Erasing data from a device or account\\
- General privacy topics (e.g., Network privacy settings; Camera privacy; Mic privacy; Physical privacy; Device privacy)\\
- Comparisons between competing privacy options (e.g., Single sign on vs. new account), or advice on which application is more private.\\
The following posts are not related to this category:\\
- If the post only discusses account security or passwords.\\
- If the post only discusses how a person's privacy has been violated, or a worry about privacy without mention of any specific privacy tools.\\
- Posts where we have to assume the use of a privacy tool since it isn't explicitly mentioned.\\
\textbf{Category-5: Compromise}\\
\textit{Definition}: Mentions of an account or device being compromised by an (implied or explicitly noted) attacker.\\
Posts related to this category may discuss any of the following:\\
- Account hacking, theft, or compromise, including changing account details\\
- Device hacking, theft, or compromise\\
- Suspicious sign-in attempt or account activity\\
- Suspicious verification message or OTP\\
- Suspicious device behavior or suspected virus\\
- Concern whether an account or device is at-risk of compromise\\
- Malware, adware, viruses, spyware, ransomware, or unwanted software\\
- Risky downloads or suspicious files\\
\textbf{Category-6: Platform actions}
\textit{Definition}: Mentions of a safety control or policy, typically by a platform or third party (such as an employer), that has gone awry and/or is causing friction with the user.\\
Posts related to this category may discuss any of the following:\\
- Person violated a platforms policy\\
- Account was banned, blocked, suspended, removed, or deleted by a platform\\
- Account was frozen or disabled by a platform\\
- IP address was blocked\\
- Access to a website or service was blocked, removed, revoked, or restricted\\
- Functionality for a service was inappropriately disabled\\
- Content being hidden, taken down, or deleted due to rule violation\\
- Payment was prevented or incorrectly flagged as fraud\\
- Person was falsely reported to a platform\\
- Foreign payment was blocked or restricted, or transaction incorrectly flagged as fraud\\
- Service is banned or prohibited in a region (due to export controls)\\
The following posts are not related to this category:\\
- Discusses being unable to login to their account for a reason other than being blocked\\
\textbf{Category-7: Scams}
\textit{Definition}: Mentions of fraud, scams, or impersonation for the apparent purposes of a scam or fraud. Fraud is defined: ``An act committed by predatory others and involves an act of deception using a false promise or a threat; a broad category where `scam' is a subset.'' Scam is defined: ``The intentional use of deceit, a trick or some dishonest means to deprive another of their money or property; a subset of fraud.''\\
Posts related to this category may discuss any of the following:\\
- Being the victim or target of a scam, fraud, sextortion, crypto scam, fake job posting, or employment scams\\
- Scams using a person's personal information\\
- Bots, spam, and unwanted communication\\
- Unauthorized or unexplained charges or transactions\\
- Stolen or missing cryptocurrency funds\\
- Stolen or missing credit card funds\\
- Problems or delays with a transaction, including cryptocurrency funds\\
- Other fraud such as incomplete purchases, blocked transactions, stolen personal details, missing goods, and lost money\\
- Fake accounts, suspicious websites, suspicious URLs, suspicious messages, or suspicious receipts\\
- Browser warnings for an unsafe webpage (e.g., phishing)\\
- Trustworthyness of a website\\
- Impersonation or identity theft\\
The following posts are not related to this category:\\
- Bots, spam, and unwanted communications for the purpose of harassment\\
- Discussing an account or device being hacked\\
\textbf{Category-8: Harassment}\\
\textit{Definition}: Interpersonal harassment or abuse by an attacker.\\
Posts related to this category may discuss any of the following:\\
- Non-consensual explicit imagery, Child sexual abuse, grooming, Cyberflashing, unwanted nude images, or image-based sexual harassment\\
- Interpersonal abuse or Intimate partner violence, including evidence collection for the same\\
- Impersonation for the purposes of harassment or abuse\\
- Stalking, monitoring \& controlling an individual through surveillance or monitoring devices\\
- Requesting or controlling victim's devices and communication channels, or accessing device without consent\\
- Harassment via text messages and phone calls, doxxing, intentionally exposing private info, threats of violence, toxic content, trolling, bullying\\
\textbf{Category-9: Data concerns}\\
\textit{Definition}: The user wants to take an action involving their data (such as hiding, reducing visibility, deletion, etc.), or is concerned or uneasy about the way their data is being used by or exposed to others (typically a platform).\\
Posts related to this category may discuss any of the following:\\
- Data collection concerns such as consent, Data retention, Data overcollection, Data harvesting, terms of service.\\
- Data sharing concerns such as with third parties, sharing without consent, unintended sharing\\
- Privacy concerns such as if a mic is listening, a camera is watching, a device always listening, or if the device is being scanned\\
- Privacy concerns such as if someone can find the user\\
- Data loss risk\\
- Data visibility, exposure or leak concerns such as unintended leaks, data breaches, password leaks, identity leaks\\
- Data retention\\
- Data misuse including unexplained traffic, ads, or a concern for how a person's data is being used\\
- Unexplained app behavior where content wont disappear, or an app is not respecting privacy settings\\
The following posts are not related to this category:\\
- Data loss risk in the presence of an attacker\\
- Post discusses specific privacy controls\\
- Post discusses sextortion or nude images, including unintended leakage\\
- Post discusses scams or fraud\\
-----------------------------------------

For a given Reddit post and a topical category, you are to identify if the topical category is applicable to the post by providing your answer as a `Yes' or `No' value.\\

Now provide your decision for the following Reddit post.\\
$<$Post$>$\\
\{\{ text \}\}\\
$<$/Post$>$\\
$<$Subreddit$>$\\
\{\{ subreddit \}\}\\
$<$/Subreddit$>$\\
$<$Topical Category$>$\\
\{\{ category \}\}\\
$<$/Topical Category$>$\\
$<$Answer$>$\\
}
  \end{tcolorbox}

\end{document}